\documentclass[%
 reprint,
 onecolumn,
 amsmath,amssymb,
 aps,
 prb,
 longbibliography,
 superscriptaddress,
]{revtex4-2}

\usepackage{graphicx}
\usepackage{dcolumn}
\usepackage{bm}
\usepackage{mathrsfs}
\usepackage[dvipsnames]{xcolor}
\usepackage{mathrsfs}
\usepackage[colorlinks=true,linkcolor=blue,citecolor=blue,urlcolor=blue,bookmarks=False]{hyperref} 
\usepackage{braket}
\usepackage{pdfpages} 
\usepackage{pgffor} 
\usepackage{tikz} 
\usetikzlibrary{arrows.meta,shapes.geometric,positioning} 

\usepackage{xcolor}
\usepackage{listings}
\usepackage[most]{tcolorbox}

\definecolor{headergray}{gray}{0.8}
\definecolor{backgray}{gray}{0.97}
\definecolor{codegray}{gray}{0.4}

\lstdefinestyle{myboxstyle}{
    basicstyle=\ttfamily,
    commentstyle=\itshape\color{codegray},
    keywordstyle=\bfseries\color{black},
    breaklines=true,
    numbers=none,
    showspaces=false,
    showstringspaces=false,
    tabsize=2,
    columns=fullflexible,
    morecomment=[l]{\#} 
}

\newtcblisting[auto counter]{mylisting}[1][]{
    width=0.7\textwidth,
    center,
    colback=backgray,
    colframe=headergray,
    coltitle=black,
    fonttitle=\bfseries,
    title=Listing,
    arc=2pt,
    boxrule=1pt,
    drop shadow,
    listing only,
    listing options={style=myboxstyle}, 
    #1
}

\makeatletter
\AtBeginDocument{\let\LS@rot\@undefined}
\makeatother

\newif\ifarXiv
\arXivfalse

\begin{document}

\preprint{APS/123-QED}

\title{LetzElPhC: an electron-phonon code for efficient exciton-phonon workflows}

\author{Muralidhar Nalabothula}
\email{muralidharrsvm7@gmail.com}
\affiliation{Department of Physics and Materials Science, University of Luxembourg, L-1511 Luxembourg, Luxembourg.}
\author{Ludger Wirtz}%
\affiliation{Department of Physics and Materials Science, University of Luxembourg, L-1511 Luxembourg, Luxembourg.}

\date{\today}

\begin{abstract}
We present \texttt{LetzElPhC}, an open-source software package that computes electron-phonon matrix elements from the output of density functional perturbation theory calculations performed with the \texttt{Quantum ESPRESSO} code. By interfacing directly with the \texttt{Yambo} code and exploiting crystal symmetries, \texttt{LetzElPhC} enables a scalable workflow for evaluating exciton-phonon matrix elements while enforcing strict gauge consistency between one-electron Bloch states and Bethe-Salpeter excitonic wave functions across the full Brillouin zone. This symmetry-preserving approach avoids phase ambiguities and significantly reduces both computational costs and memory requirements. To support large-scale simulations on dense momentum grids, the package incorporates Fourier interpolation of dynamical matrices and perturbed Kohn-Sham potentials, accounting for long-range dipole and quadrupole interactions in both two- and three-dimensional materials, combined with multi-level parallelization and efficient parallel I/O. Together, these capabilities seamlessly complement the \texttt{Yambo} code to enable first-principles simulations of phonon-assisted optical spectroscopies, which we demonstrate through applications to resonant Raman scattering in monolayer $\text{WSe}_2$ and phonon-assisted luminescence in cubic AgBr.
\end{abstract}

 \maketitle

\section{Introduction}
\label{sec:intro}

The interaction between electrons and phonons (lattice vibrations) underlies a wide range of phenomena in condensed matter physics, including conventional superconductivity~\cite{PhysRev.108.1175}, charge-density waves~\cite{Gruener1987,10.1098/rspa.1954.0116}, and carrier transport properties~\cite{PhysRev.80.72,Ponce2020Feb}. Electron-phonon interactions also renormalize electronic states, govern their temperature dependence, and determine quasiparticle lifetimes~\cite{RevModPhys.89.015003,grimvall1981electron}. Beyond these quasiparticle effects, they play an essential role in excited-state phenomena, where lattice vibrations couple to collective electronic excitations.

In particular, there has been considerable recent interest in the role of electron-phonon interactions in many-body optical excitations, described in terms of the coupling between excitons (bound electron-hole pairs) and phonons~\cite{PhysRevB.105.085111,Chan2023May,PhysRevLett.125.107401,PhysRevB.99.081109,PhysRevLett.122.187401,Reichardt2020Aug,Marini2024May,Nalabothula2025Apr,PhysRevB.106.125403}. In two-dimensional materials such as transition metal dichalcogenides (TMDCs)~\cite{Chan2023May,Marini2024May}, wide-band-gap semiconductors such as hexagonal boron nitride ($h$BN)~\cite{PhysRevLett.131.206902,PhysRevLett.122.187401}, and magnetic semiconductors~\cite{PhysRevB.110.155131}, exciton-phonon interactions are essential for accurately describing exciton dynamics, determining exciton relaxation times~\cite{Chan2023May,PhysRevLett.125.107401,PhysRevResearch.4.043203}, and understanding optical processes such as phonon-assisted luminescence~\cite{Cassabois2016Apr,PhysRevB.99.081109,PhysRevLett.122.187401,Marini2024May} and resonant Raman scattering~\cite{Yu2010,Reichardt2020Aug,Nalabothula2025Apr}.

From a computational perspective, the evaluation of exciton-phonon coupling requires four key ingredients: one-electron Bloch states, dynamical matrices, the change in the Kohn-Sham potential due to atomic displacements, and excitonic states. The starting point is the calculation of one-electron Kohn-Sham wave functions and energies within density functional theory (DFT). Phonon dynamical matrices, together with the change in the self-consistent potential due to atomic displacements, are then obtained using density functional perturbation theory (DFPT)~\cite{RevModPhys.73.515}, as implemented in plane-wave codes such as \texttt{Quantum ESPRESSO}~\cite{Giannozzi2017Oct} and \texttt{Abinit}~\cite{Verstraete2025Oct}, or through finite-difference approaches~\cite{Poliukhin2026Mar,Monserrat2018Feb,PhysRevB.100.174304}. Electron-phonon matrix elements are subsequently evaluated as transition amplitudes of the variation in the self-consistent potential between initial and final Kohn-Sham states. Finally, excitonic energies and wave functions are obtained by solving the Bethe-Salpeter equation (BSE)~\cite{PhysRevB.21.4656,PhysRevLett.81.2312,PhysRevLett.80.4510,PhysRevLett.80.4514}, typically on top of a GW-corrected~\cite{PhysRevB.34.5390} DFT electronic structure. Several modern open-source \textit{ab initio} many-body codes provide BSE implementations, including \texttt{Yambo}~\cite{Marini2009Aug,Sangalli2019May}, \texttt{BerkeleyGW}~\cite{Deslippe2012Jun}, \texttt{WEST}~\cite{Govoni2015Jun}, \texttt{Abinit}~\cite{Verstraete2025Oct}, \texttt{Exciting}~\cite{Gulans2014Aug}, and \texttt{GPAW}~\cite{Mortensen2024Mar}. These ingredients are then combined to evaluate exciton-phonon matrix elements~\cite{PhysRevB.105.085111,PhysRevLett.125.107401,Reichardt2020Aug}.

Despite the exciton-phonon workflow being conceptually straightforward as a post-processing step, particularly with the availability of electron-phonon packages such as \texttt{EPW}~\cite{Lee2023Aug}, \texttt{EPIq}~\cite{Marini2024Feb}, \texttt{Perturbo}~\cite{Zhou2021Jul}, and \texttt{Phoebe}~\cite{Cepellotti2022Aug}, an important caveat remains. The Kohn-Sham wave functions used to solve the BSE must exactly match those used to evaluate the electron-phonon matrix elements. This requirement ensures that the resulting exciton-phonon matrix elements remain invariant under arbitrary phase choices of the one-electron Bloch states.

In practice, maintaining gauge consistency across the full workflow is challenging, particularly when combining electron-phonon quantities with BSE calculations performed by different codebases. To avoid inconsistencies arising from phase mismatches, existing workflows often disable crystal symmetries during the computation of electron-phonon quantities. While this approach guarantees consistency, it introduces severe computational limitations in subsequent many-body calculations, where  symmetry-based reductions are essential to control computational cost and memory usage~\cite{7h2w-ys9h}. As a result, calculations involving exciton-phonon interactions quickly become prohibitively expensive, severely limiting accessible system sizes and material complexity.

These challenges can be further amplified by the fact that converged BSE calculations typically require very fine Brillouin-zone samplings, leading to large electron-phonon databases. Efficient handling of these datasets therefore requires modern high-performance computing techniques, including portable parallel I/O and scalable parallelization strategies capable of distributing electron-phonon calculations over large numbers of CPUs. In addition, DFPT calculations can be computationally demanding and, in some cases, prohibitive, making interpolation schemes for dynamical matrices and variations of the Kohn-Sham potential essential for large-scale simulations.
Beyond computational efficiency, it is also important that such workflows remain compatible with a broad range of electronic structure systems, including magnetic systems, where excitonic effects can be particularly strong~\cite{PhysRevB.110.155131}. However, support for these systems remains limited in many existing implementations, thereby restricting the applicability of exciton-phonon calculations in magnetic semiconductors.

In this paper, we present \texttt{LetzElPhC} (\emph{L\"etzebuerg Electron-Phonon Code}, where \emph{L\"etzebuerg} is Luxembourgish for Luxembourg), an open-source, high-performance, and portable software package written in C, which is publicly available on GitHub~\cite{letzcode_github}. \texttt{LetzElPhC} computes electron-phonon matrix elements within the norm-conserving pseudopotential framework by combining DFPT outputs from \texttt{Quantum ESPRESSO} with wave function data stored in \texttt{Yambo} databases, thereby ensuring direct compatibility with subsequent many-body calculations within the \texttt{Yambo} code. This design enables the consistent use of crystal symmetries throughout the workflow and enforces gauge consistency between electron-phonon and BSE wave functions, thereby avoiding phase-related ambiguities in exciton-phonon calculations. \texttt{LetzElPhC} further exploits crystal symmetries, supports multiple levels of parallelization, and provides portable parallel I/O for efficient and scalable data handling. The code is designed to scale from laptops and workstations to large-scale HPC clusters. It supports a broad class of electronic structure systems, including magnetic systems and two- and three-dimensional materials, as well as interpolation schemes for phonon-related quantities. As a result, \texttt{LetzElPhC} enables efficient and scalable exciton-phonon calculations with reduced computational cost while remaining applicable to complex material systems.

\section{Methodology}
\subsection{Electron-phonon matrix elements}
\label{sec:elph}
The central quantity evaluated by the \texttt{LetzElPhC} program is the electron-phonon matrix element, which gives the transition amplitude of an electron scattering between two Bloch states due to the perturbation introduced by a specific lattice vibration. Therefore, we begin with the expression for the electron-phonon matrix element, which is given by:
\begin{equation}
    \mathfrak{g}^\nu_{mn}(\mathbf{k},\mathbf{q}) = \langle m, \mathbf{k}+\mathbf{q} \mid \partial^\nu_{\mathbf{q}} V \mid n, \mathbf{k} \rangle,\label{eq:gkkp}
\end{equation}
where $\mathbf{k}$ and $\mathbf{q}$ are the crystal momenta of the electron and the phonon, respectively, and $|n,\mathbf{k}\rangle$ represents the one-electron Bloch state of an electron with band index $n$ and crystal momentum $\mathbf{k}$. The operator $\partial^{\nu}_{\mathbf{q}} V$ denotes the directional derivative of the self-consistent potential along the displacement vector (with dimensions of length) of the phonon mode with index $\nu$ and is written as:

\begin{equation}
    \partial^{\nu}_{\mathbf{q}} V_{\sigma \sigma'} = \sqrt{\frac{\hbar}{2\omega^{\nu}_\mathbf{q}}} \sum_{l\kappa\alpha} e^{i\mathbf{q}\cdot\mathbf{R}_l}\frac{(\mathbf{d}^{\nu}_{\mathbf{q}})_{\kappa\alpha}}{\sqrt{M_{\kappa}}}
    \frac{\partial V_{\sigma \sigma'}}{\partial R_{l\kappa\alpha}} ,\label{eq:dvq_def}
\end{equation}
where $M_{\kappa}$ is the mass of the atom $\kappa$, located at position $\mathbf{R}_{l\kappa}$ within the $l$-th unit cell positioned at $\mathbf{R}_{l}$, $(\mathbf{d}^{\nu}_{\mathbf{q}})_{\kappa\alpha}$ is the $\nu$-th eigenvector of the phonon dynamical matrix at crystal momentum $\mathbf{q}$, $\alpha$ denotes the Cartesian direction, $\omega_\mathbf{q}^{\nu}$ is the frequency of the $\nu$-th phonon mode, and $\sigma, \sigma'$ are the spin indices. The operator $\partial^{\nu}_{\mathbf{q}} V_{\sigma \sigma'}$ can be written in terms of a lattice-periodic function, i.e.,
\begin{equation}
    \partial^{\nu}_{\mathbf{q}} V_{\sigma \sigma'} (\mathbf{r}) = e^{i\mathbf{q}\cdot\mathbf{r}} \partial^{\nu}_{\mathbf{q}} v_{\sigma \sigma'}(\mathbf{r}),
    \label{eq:dvscf_perdioc}
\end{equation}
where $\partial^{\nu}_{\mathbf{q}} v_{\sigma \sigma'}(\mathbf{r})$ is a lattice-periodic function of real space, satisfying
$\partial^{\nu}_{\mathbf{q}} v_{\sigma \sigma'}(\mathbf{r})=\partial^{\nu}_{\mathbf{q}} v_{\sigma \sigma'}(\mathbf{r}+\mathbf{T})$
for any lattice translation vector $\mathbf{T}$. The full perturbing potential $\partial^{\nu}_{\mathbf{q}} V_{\sigma \sigma'}(\mathbf{r})$, on the other hand, is periodic in reciprocal space, i.e.,
$\partial^{\nu}_{\mathbf{q}} V_{\sigma \sigma'}(\mathbf{r})=\partial^{\nu}_{\mathbf{q}+\mathbf{G}} V_{\sigma \sigma'}(\mathbf{r})$,
where $\mathbf{G}$ is a reciprocal lattice vector.

Within the norm-conserving pseudopotential framework~\cite{PhysRevLett.43.1494,PhysRevLett.48.1425,PhysRevB.43.1993,PhysRevB.40.2980}, the one-electron Kohn-Sham states $\ket{n, \mathbf{k}}$ are replaced with the pseudo one-electron states $\ket{\widetilde{n, \mathbf{k}}}$, and the Kohn-Sham potential is replaced with the pseudo-Kohn-Sham potential ($V^{\text{PS}}$), which is given by 
\begin{equation}
     V^{\text{PS}}_{\sigma \sigma'}(\mathbf{r},\mathbf{r'}) = V^{\text{local}}_{\sigma \sigma'}(\mathbf{r})
    \delta(\mathbf{r}-\mathbf{r'}) +  V^{\text{NL}}_{\sigma \sigma'}(\mathbf{r},\mathbf{r'}).
    \label{eq:kd_pseudo_pot}
\end{equation}

Here $V^{\text{local}}_{\sigma \sigma'}(\mathbf{r})$ is the local potential, which is the sum of the electron-ion, Hartree, and exchange-correlation potentials, and $V^{\text{NL}}_{\sigma \sigma'}(\mathbf{r},\mathbf{r'})$ is the non-local potential (see Ref.~\cite{PhysRevLett.48.1425} for the explicit form). In \texttt{LetzElPhC}, the Hartree and exchange-correlation contributions to the variation of the local potential are read directly from the DFPT outputs of \texttt{Quantum ESPRESSO}, while the local ionic potential and non-local pseudopotential contributions are computed internally using the formalism described in Refs.~\cite{PhysRevB.64.235118,PhysRevB.71.115106}.
\begin{figure}
    \centering
    \includegraphics[width=0.5\linewidth]{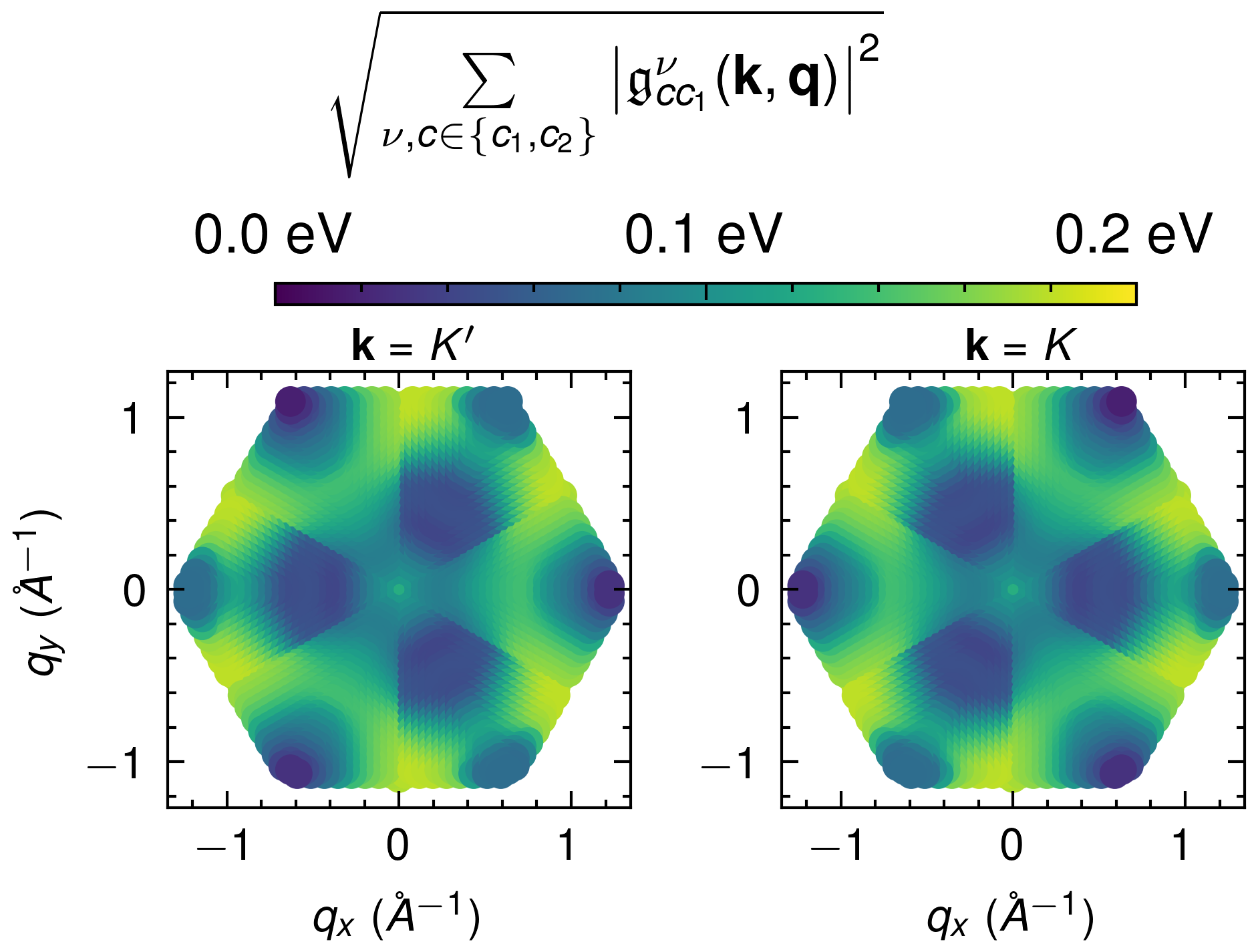}
    \caption{Gauge-invariant electron-phonon coupling strength from the lowest conduction-band states at the $K/K'$ valleys in monolayer WSe$_2$ to the two lowest conduction bands, summed over all phonon modes as a function of phonon crystal momentum $\mathbf{q}$.}
    \label{fig:wse2_elph}
\end{figure}

As an example, we show in Fig.~\ref{fig:wse2_elph} the electron-phonon scattering coupling in monolayer WSe$_2$ as a function of phonon crystal momentum $\mathbf{q}$. We consider an electron in the lowest conduction band at $\mathbf{k} = K\equiv(1/3,1/3,0)$ or $\mathbf{k} = K'\equiv(-1/3,-1/3,0)$ scattering into the two lowest conduction band states at all other crystal momenta in the Brillouin zone, summed over all phonon modes. The coupling strength exhibits a three-fold rotational symmetry and is dominated by $M$-point phonons, consistent with previous work on monolayer MoS$_2$~\cite{Chan2023May}. The dominant process involves scattering of electronic states from $K \rightarrow \Gamma K'/2$ and $K' \rightarrow \Gamma K/2$ via $M$-point phonons. The dominance of $M$-point phonons is attributed to the van Hove singularity in the phonon dispersion at $M$~\cite{huang2022first,guo2015double,ty8m-mgml}.
\subsection{Exciton-Phonon Coupling}
\label{sec:exph_coupling}
Although \texttt{LetzElPhC} is designed to compute electron-phonon matrix
elements, its primary purpose is to provide the electron-phonon coupling
quantities required for efficient exciton-phonon calculations. The evaluation
of exciton-phonon observables is instead handled separately in Python scripts, which are also provided with the code~\cite{nalabothula_2026_24130523}. This separation
ensures modularity and facilitates the rapid development of new
exciton-phonon functionalities while keeping the core electron-phonon
calculation independent of specific observables. We briefly summarize the
formalism here, as it highlights the technical challenges that arise in the
practical implementation of exciton-phonon workflows.

The exciton-phonon matrix element
$\mathscr{G}_{S'S}^\lambda(\mathbf{Q},\mathbf{q})$ describes the scattering
of an exciton from an initial state $|S,\mathbf{Q}\rangle$ to a final state
$|S',\mathbf{Q}+\mathbf{q}\rangle$ mediated by a phonon of branch
$\lambda$ and wavevector $\mathbf{q}$. Excitonic states obtained by solving the BSE can be expressed in the basis of
electron-hole transitions as~\cite{PhysRevB.21.4656}
\begin{equation}
    |S,\mathbf{Q}\rangle =
    \sum_{c,v,\mathbf{k}}
    X_{\mathbf{k}cv}^{S;(\mathbf{Q})}
    \hat{a}_{c\mathbf{k}}^\dagger
    \hat{a}_{v\mathbf{k}-\mathbf{Q}}
    |0\rangle ,
    \label{eq:exwf}
\end{equation}
where $\hat{a}^{\dagger}$ and $\hat{a}$ are electron creation and
annihilation operators, respectively, and
$X_{\mathbf{k}cv}^{S;(\mathbf{Q})}$ are the exciton wave function
coefficients in the electron-hole basis. Using these coefficients and the electron-phonon matrix elements
$\mathfrak{g}_{mn}^{\lambda}(\mathbf{k},\mathbf{q})$, the exciton-phonon matrix element can be written as~\cite{7h2w-ys9h,PhysRevB.105.085111,PhysRevLett.125.107401,PhysRevLett.131.206902}
\begin{equation}
\begin{aligned}
    \mathscr{G}^{\lambda}_{S'S}(\mathbf{Q},\mathbf{q}) =\;
    &\sum_{\mathbf{k},c,c',v}
      \bigl(X^{S';(\mathbf{Q}+\mathbf{q})}_{\mathbf{k}+\mathbf{q}c'v}\bigr)^*
      X^{S;(\mathbf{Q})}_{\mathbf{k}cv}\,
      \mathfrak{g}^{\lambda}_{c'c}(\mathbf{k},\mathbf{q}) \\
    -\;&\sum_{\mathbf{k},c,v,v'}
      \bigl(X^{S';(\mathbf{Q}+\mathbf{q})}_{\mathbf{k}cv'}\bigr)^*
      X^{S;(\mathbf{Q})}_{\mathbf{k}cv}\,
      \mathfrak{g}^{\lambda}_{vv'}(\mathbf{k}-\mathbf{Q}-\mathbf{q},\mathbf{q}) .
\end{aligned}
    \label{eq:exph_matrix}
\end{equation}
The first term corresponds to electron scattering between conduction bands $c$ and $c'$, whereas the second term describes hole scattering between valence bands $v$ and $v'$.

Equation~\eqref{eq:exph_matrix} immediately reveals the stringent requirements
that a practical exciton-phonon workflow must satisfy. The exciton coefficients
$X_{\mathbf{k}cv}^{S;(\mathbf{Q})}$ inherit the arbitrary phases of the
underlying Bloch states, while the electron-phonon matrix elements
$\mathfrak{g}_{mn}^{\lambda}(\mathbf{k},\mathbf{q})$ carry the same phase
dependence. Therefore, the exciton-phonon matrix elements
$\mathscr{G}^{\lambda}_{S'S}$ in Eq.~\eqref{eq:exph_matrix} are independent of
the arbitrary Bloch-state phases only when the excitonic states and
electron-phonon matrix elements are evaluated using the same set of Bloch
states. A straightforward approach to avoid these phase inconsistencies is to perform
the entire exciton-phonon workflow using the same one-electron wave functions
and excitonic states throughout the full Brillouin zone (FBZ), without
exploiting crystal symmetries. However, the convergence of excitonic states
requires very dense $\mathbf{k}$-point sampling, making such an
approach without crystal symmetries computationally very expensive.

In addition to the challenges associated with phase and symmetry consistency,
exciton-phonon calculations require dense phonon $\mathbf{q}$-point meshes to converge
properties such as scattering rates and lifetimes~\cite{PhysRevLett.125.107401,PhysRevResearch.4.043203,Chan2023May}.
These requirements lead to momentum grids that are significantly denser than
those accessible through direct DFPT calculations. Therefore, exploiting
crystal symmetries and efficiently interpolating the Kohn-Sham potential
variations onto arbitrarily fine momentum grids are essential for large-scale
exciton-phonon calculations.

\begin{figure}
    \centering
    \includegraphics[width=0.35\linewidth]{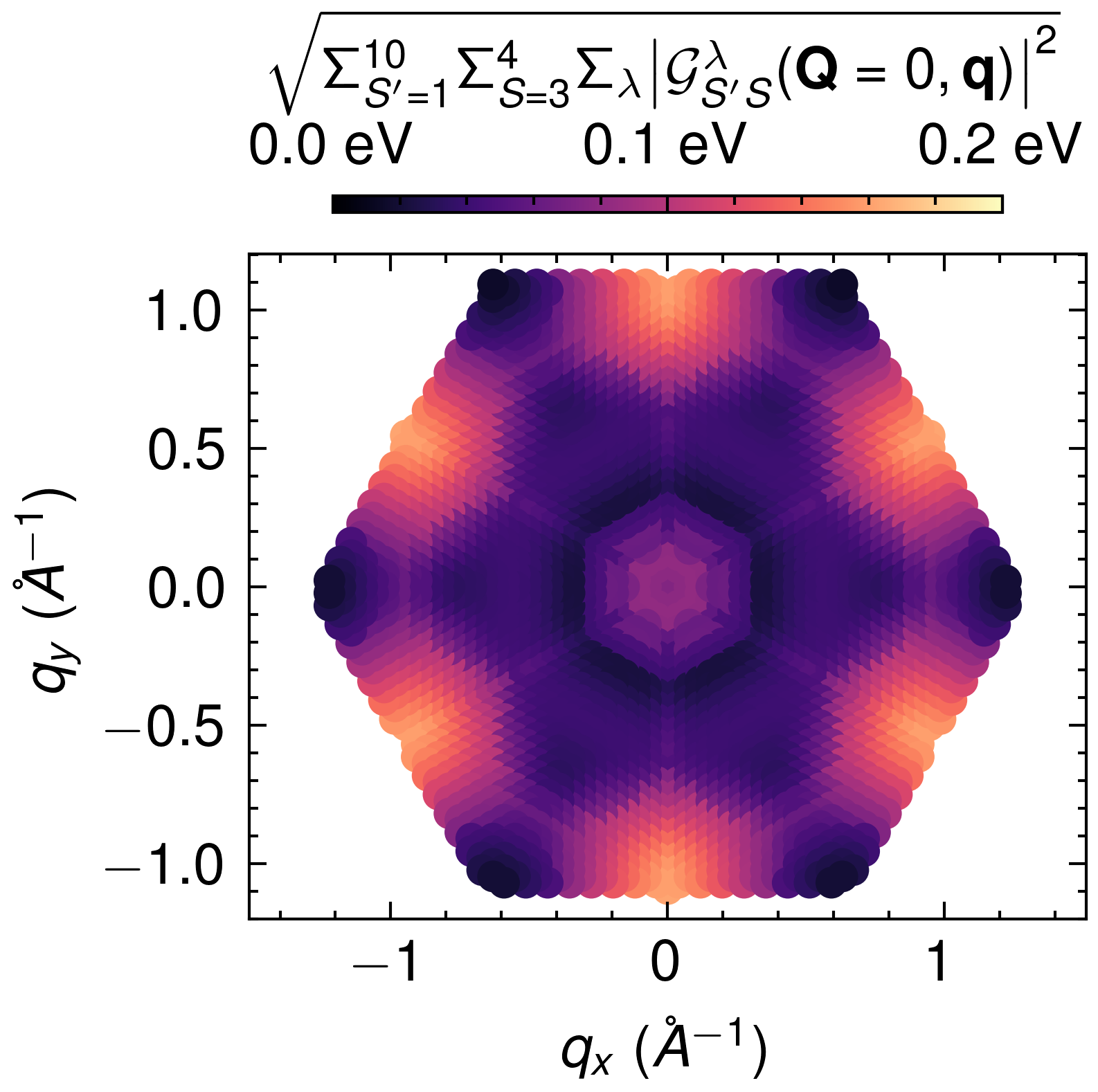}
    \caption{Gauge-invariant exciton-phonon coupling strength from the two degenerate
lowest in-plane bright excitons at the $\Gamma$ point (the third- and
fourth-lowest excitonic states in energy) to the 10 lowest-energy finite crystal momentum excitonic
states in monolayer WSe$_2$, summed over all
phonon modes.}
    \label{fig:wse2_exph}
\end{figure}

To validate our approach, we apply our exciton-phonon workflow to monolayer WSe$_2$. In Fig.~\ref{fig:wse2_exph}, we show the exciton-phonon coupling strength of the lowest doubly degenerate in-plane bright excitons with zero crystal momentum scattering to the ten lowest exciton states at different crystal momenta throughout the Brillouin zone, considering all phonon modes. The exciton-phonon coupling strength respects the three-fold rotational symmetry of the crystal, and the dominant scattering pathway involves bright excitons scattering to the $M$-point excitons via $M$-point phonons, consistent with previous work on monolayer MoS$_2$~\cite{Chan2023May}. Given that the lowest $M$-point excitons are mostly composed of a hole at the $K/K'$ point and an electron at $-Q/Q$~\cite{Chan2023May} ($Q = \frac{\Gamma K}{2}$), this implies that the exciton-phonon scattering is predominantly due to electron scattering from $K/K'$ to $-Q/Q$, which can be attributed to the dominant electron-phonon coupling from the $M$-point phonons, as shown in Fig.~\ref{fig:wse2_elph}.

\subsection{Symmetries}
Evaluating electron-phonon matrix elements on the dense momentum grids required
for converged many-body calculations is computationally demanding. The cost can
be reduced substantially by restricting calculations to the irreducible
Brillouin zone (IBZ) and expanding to the full Brillouin zone (FBZ) via crystal
symmetries. However, a naive application of this procedure breaks the gauge
consistency required by BSE solvers. Therefore, \texttt{LetzElPhC} implements a symmetry-unfolding procedure to restore this consistency.

In this subsection, we briefly outline how crystal symmetries are used to obtain
wave functions, perturbed Kohn-Sham potentials, and phonon eigenvectors on the
FBZ from their IBZ counterparts, and how the electron-phonon matrix elements
can be constructed throughout the entire Brillouin zone from a minimal set of
explicitly computed matrix elements (see Refs.~\cite{7h2w-ys9h,RevModPhys.40.1,PhysRevB.76.165108} for detailed derivations).

Consider a spatial symmetry operation $g = \{R \mid \bm{\tau}\} \in \mathcal{G}$, 
where $\mathcal{G}$ denotes the space group of the crystal. Under the action of $g$, 
a Bloch wave function $\phi_{\mathbf{k},m}(\mathbf{r})$ with crystal momentum 
$\mathbf{k}$ and band index $m$ transforms as~\cite{altmann2005rotations,
el2008symmetry}
\begin{equation}
\left( \{R \mid \bm{\tau}\} \ \phi_{\mathbf{k},m} \right)_{\sigma} (\mathbf{r})
\equiv
\sum_{\sigma'}
S_{\sigma\sigma'}(R)\,
\left(\phi_{\mathbf{k},m}\right)_{\sigma'}
\!\left(R^{-1}(\mathbf{r}-\bm{\tau})\right),
\end{equation}
where $S(R)$ is a $2\times2$ unitary matrix acting on the spin indices $\sigma$ and 
$\sigma'$. In the absence of spin, $S(R)$ reduces to the $1\times1$ identity matrix. 
The transformed wave function corresponds to a Bloch state at crystal momentum 
$R\mathbf{k}$. In general, it may differ from the actual wave function by a trivial 
phase factor or, in the presence of degeneracies, by a unitary rotation. It can 
therefore be expressed as
\begin{equation}
\left( \{R \mid \bm{\tau}\} \  \phi_{\mathbf{k},m} \right)(\mathbf{r})
=
\sum_{m'}
\mathcal{D}_{\mathbf{k},m'm}(g)\,
\phi_{R\mathbf{k},m'}(\mathbf{r}),
\label{eq:psi_rot}
\end{equation}
where $\mathcal{D}_{\mathbf{k}}(g)$ is a unitary matrix. Assuming that the Bloch 
states form an orthonormal basis, the matrices $\mathcal{D}_{\mathbf{k}}(g)$ are 
given by~\cite{Iraola2022Mar,Matsugatani2021Jul}
\begin{equation}
\mathcal{D}_{\mathbf{k},m'm}(g)
=
\sum_{\sigma,\sigma'}
\int d^3\mathbf{r}\,
\left(\phi^*_{R\mathbf{k},m'}\right)_{\sigma'}(\mathbf{r})
\,S_{\sigma'\sigma}(R)\,
\left(\phi_{\mathbf{k},m}\right)_{\sigma}
\!\left(R^{-1}(\mathbf{r}-\bm{\tau})\right).
\label{eq:Dmat}
\end{equation}
Throughout this work, we adopt the periodic gauge convention, i.e., 
$\phi_{\mathbf{k},m}(\mathbf{r})=\phi_{\mathbf{k}+\mathbf{G},m}(\mathbf{r})$, 
where $\mathbf{G}$ is a reciprocal lattice vector.

Using Eq.~\eqref{eq:psi_rot}, the wave functions in the FBZ can be obtained by applying symmetry operations to the wave functions computed in the IBZ and setting the corresponding $\mathcal{D}_{\mathbf{k}}(g)$ matrices to the identity. It should 
be noted that the symmetry operation mapping $\mathbf{k}$ to $R\mathbf{k}$ ($R\mathbf{k} \neq \mathbf{k} + \mathbf{G}$) is not unique. Consequently, fixing a particular choice of $\mathcal{D}_{\mathbf{k}}(g)$ defines a gauge convention that must be maintained consistently throughout the calculation.

Similar symmetry relations apply to phonon eigenvectors and the variation of the 
Kohn-Sham potential. Therefore, these quantities can also be computed solely within the IBZ and subsequently mapped to the FBZ using symmetry 
operations~\cite{RevModPhys.40.1,PhysRevB.76.165108}. For phonons, the action of a 
symmetry operation on a phonon eigenvector $\mathbf{d}_{\mathbf{q}}^\lambda$ of mode 
index $\lambda$ and crystal momentum $\mathbf{q}$ yields the phonon eigenvector at $R\mathbf{q}$, i.e.,
\begin{equation}
\left ( \{R \mid \bm{\tau}\} \ \mathbf{d}^\lambda_{\mathbf{q}} \right)_{\tilde{\kappa}\alpha}
\equiv
\sum_{\kappa,\beta}
e^{i \mathbf{q} \cdot (g^{-1} \boldsymbol{\tau}_{\tilde{\kappa}} - 
\boldsymbol{\tau}_{\kappa})}
R_{\alpha \beta}
d^\lambda_{\mathbf{q},\kappa\beta} \delta_{\tilde{\kappa},g(\kappa)}
=
\sum_{\lambda'}
\Gamma_{\mathbf{q},\lambda'\lambda}(g)
d^{\lambda'}_{R\mathbf{q},\tilde{\kappa}\alpha},
\label{eq:ph_rot}
\end{equation}
where $\Gamma_{\mathbf{q},\lambda'\lambda}(g)$ is a unitary matrix accounting for 
the fact that the transformed eigenvector may differ from the actual eigenvector by 
a unitary rotation.

For the variation of the self-consistent Kohn-Sham potential 
$\partial_{\mathbf{q}}^\lambda V(\hat{\mathbf{r}})$, the action of the symmetry operator 
$\hat{U}(g)$ is given by~\cite{PhysRevB.76.165108,RevModPhys.40.1}
\begin{equation}
(\hat{U}(g) \partial_{\mathbf{q}}^\lambda V(\hat{\mathbf{r}}) \hat{U}(g)^{\dagger} )_{\sigma \sigma'}
\equiv
\sum_{\tilde{\sigma},\tilde{\sigma}'}
S_{\sigma\tilde{\sigma}}(R)\,
\partial_{\mathbf{q},\tilde{\sigma},\tilde{\sigma}'}^\lambda V
\left(R^{-1}(\hat{\mathbf{r}}-\bm{\tau})\right)
S^*_{\sigma'\tilde{\sigma}'}(R)
=
\sum_{\lambda'}
\Gamma_{\mathbf{q},\lambda'\lambda}(g)
\partial_{R\mathbf{q},\sigma\sigma'}^{\lambda'} V(\hat{\mathbf{r}}).
\label{eq:deformationPot}
\end{equation}

Assuming that all phonon eigenvectors are mutually orthogonal, the unitary matrices 
$\Gamma_{\mathbf{q}}(g)$ can be written as~\cite{RevModPhys.40.1}
\begin{equation}
\begin{aligned}
\Gamma_{\mathbf{q},\lambda'\lambda}(g)
&=
\left(\mathbf{d}_{R\mathbf{q}}^{\lambda'}\right)^\dagger
\left(\{R \mid \bm{\tau}\} \ \mathbf{d}^\lambda_{\mathbf{q}} \right) =
\sum_{\kappa,\beta,\tilde{\kappa},\alpha}
e^{i \mathbf{q} \cdot (g^{-1} \boldsymbol{\tau}_{\tilde{\kappa}} - 
\boldsymbol{\tau}_{\kappa})}
R_{\alpha \beta}
d^\lambda_{\mathbf{q},\kappa\beta}
\left(d^{\lambda'}_{R\mathbf{q},\tilde{\kappa}\alpha}\right)^* \delta_{\tilde{\kappa},g(\kappa)}.
\end{aligned}
\label{eq:ph_rep}
\end{equation}

Finally, by combining Eqs.~\eqref{eq:psi_rot} and \eqref{eq:deformationPot}, one 
obtains~\cite{7h2w-ys9h}:
\begin{equation}
\begin{aligned}
\mathfrak{g}^{\lambda'}_{mn}(R\mathbf{k},R\mathbf{q}) &= \sum_{m',n',\lambda} 
\Gamma^*_{\mathbf{q},\lambda'\lambda}(g)\mathcal{D}_{\mathbf{k}+\mathbf{q},mm'}(g) 
\mathcal{D}_{\mathbf{k},nn'}^*(g)\, \mathfrak{g}^{\lambda}_{m'n'}(\mathbf{k},\mathbf{q}).
\end{aligned}
\label{eq:elph_rot222}
\end{equation}

Utilizing Eq.~\eqref{eq:elph_rot222}, the electron-phonon matrix elements for any arbitrary $R\mathbf{q}$ phonon wavevector in the FBZ can be reconstructed by mapping directly from the IBZ, without explicitly evaluating the computationally expensive matrix elements given in Eq.~\eqref{eq:gkkp}. This also eliminates the need to rotate $\partial^\lambda_{\mathbf{q}} V(\mathbf{r})$ from the IBZ to the FBZ. Furthermore, when $g$ belongs to the little group of 
$\mathbf{q}$, i.e.\ $R\mathbf{q} \equiv \mathbf{q}$, these equations naturally 
yield the matrix elements at $R\mathbf{k}$ from those at $\mathbf{k}$, while 
maintaining strict gauge consistency.
The same symmetry-based approach can also be applied to excitonic
wave functions. Excitonic states throughout the FBZ can be reconstructed from
those in the IBZ using the formalism of Ref.~\cite{7h2w-ys9h}:
\begin{equation}
X^{S;(R\mathbf{Q})}_{R\mathbf{k}cv}
=
\sum_{c'v'}
\mathcal{D}_{\mathbf{k},cc'}(g)
\left(\mathcal{D}_{\mathbf{k}-\mathbf{Q},vv'}(g)\right)^*
X^{S;(\mathbf{Q})}_{\mathbf{k}c'v'}.
\label{eq:rot_exwf}
\end{equation}

Equations~\eqref{eq:elph_rot222} and~\eqref{eq:rot_exwf} allow us to compute the exciton-phonon matrix elements throughout the full Brillouin zone while fully accounting for the symmetries at every stage and without encountering any phase inconsistencies.

\subsection{Interpolation of Dynamical Matrices and Perturbed Kohn-Sham Potentials}
\label{sec:interpolation}
\begin{figure}
    \centering
    \includegraphics[width=\linewidth]{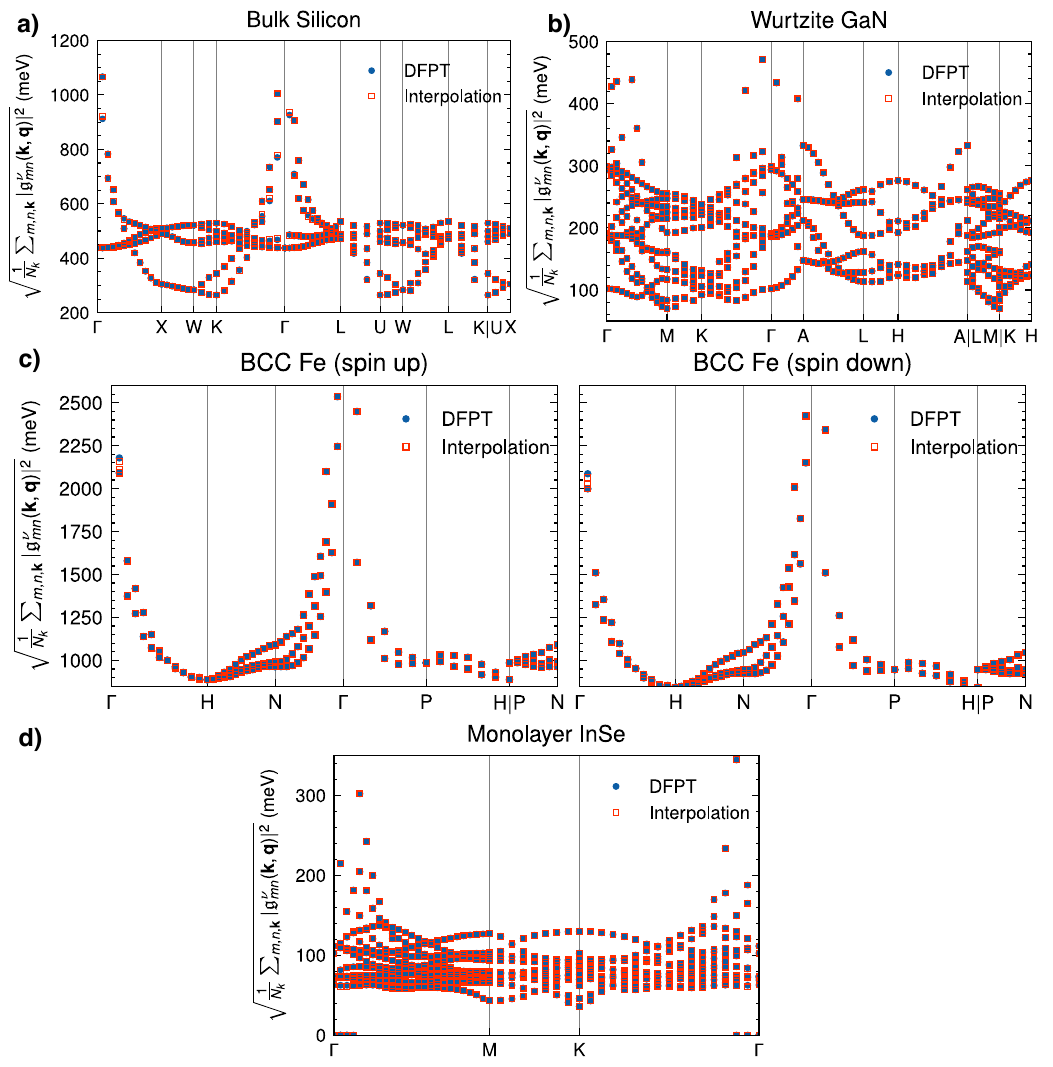}
    \caption{Brillouin-zone-averaged electron-phonon matrix elements, $\sqrt{N_{\mathbf{k}}^{-1}\sum_{m,n,\mathbf{k}} |\mathfrak{g}_{mn}^{\nu}(\mathbf{k},\mathbf{q})|^2}$, along high-symmetry phonon wavevector ($\mathbf{q}$) paths for (a)~bulk silicon, (b)~wurtzite GaN, (c)~bcc Fe for spin-up (left) and spin-down (right) channels, and (d)~monolayer InSe. Filled blue circles and open red squares represent direct DFPT calculations and interpolated values, respectively.}
    \label{fig:interpolation}
\end{figure}

\begin{table}[b]
\caption{\label{tab:interp_comp_details}DFT, DFPT, and interpolation parameters for benchmark systems: kinetic energy cutoff $E_{\text{cut}}$, coarse and fine meshes, active bands.}
\begin{ruledtabular}
\begin{tabular}{lccccc}
System & $E_{\text{cut}}$ & Coarse & Fine & Active & Long-range \\
 & (Ry) & $\mathbf{q}$-mesh & $\mathbf{k},\mathbf{q}$-mesh & bands & terms \\
\colrule
Si & 50  & $6 \times 6 \times 6$ & $24 \times 24 \times 24$ & 1--8   & Quad. \\
GaN & 120 & $8\times 8\times 6$ & $24\times 24 \times 12$ & 17--20 & Dip.+Quad. \\
Fe & 70  & $8 
\times 8 \times 8$ & $24 
\times 24 \times 24 $ & 5--12  & None \\
InSe (1L) & 120 & $12 \times 12 \times 1$ & $48\times 48\times 1$  & 28--31 & 2D Dip.+Quad. \\
\end{tabular}
\end{ruledtabular}
\end{table}

Although symmetry reduction significantly lowers the computational cost,
evaluating dynamical matrices and perturbed Kohn-Sham potentials via DFPT
for the large number of $\mathbf{q}$-points in the IBZ can be computationally demanding for complex materials. One
must therefore rely on Fourier interpolation to evaluate these quantities on
arbitrarily fine grids. Fortunately, high-quality Fourier interpolation
schemes~\cite{PhysRevB.55.10355,PhysRevB.78.045124,PhysRevLett.115.176401,
PhysRevB.92.054307} are well established for three-dimensional systems and,
more recently, for two-dimensional
materials~\cite{PhysRevB.103.075410,PhysRevB.106.115423,
PhysRevB.107.155424,PhysRevX.11.041027,PhysRevMaterials.2.114010,
Sohier2017Jun,PhysRevB.102.094308,PhysRevLett.125.136601,
PhysRevLett.125.136602,PhysRevB.102.125203}, for both dynamical matrices
and perturbed Kohn-Sham potentials. The accuracy of these schemes has been
further improved by the recent availability of codes to compute dynamical
quadrupole tensors~\cite{PhysRevX.9.021050,PhysRevB.110.094306,hc53-g1p3}.
\texttt{LetzElPhC} implements these interpolation schemes for both
three-dimensional and two-dimensional systems. In this subsection, we briefly discuss the main expressions used in this work.

To perform Fourier interpolation, the macroscopic long-range (LR) contribution 
is first subtracted from the total dynamical matrices or perturbed Kohn-Sham 
potentials~\cite{PhysRevB.55.10355,PhysRevLett.115.176401,PhysRevB.92.054307} 
to obtain the short-range (SR) component. The dynamical matrix is decomposed as
\begin{equation}
    D_{\kappa \alpha, \kappa' \alpha'}(\mathbf{q}) =
    D^{\mathrm{S}}_{\kappa \alpha, \kappa' \alpha'}(\mathbf{q}) +
    D^{\mathrm{L}}_{\kappa \alpha, \kappa' \alpha'}(\mathbf{q}),
    \label{eq:Dyn_LR_SR}
\end{equation}
where the superscripts $\mathrm{S}$ and $\mathrm{L}$ denote the short-range
and long-range contributions, respectively. Similarly, the perturbed
Kohn-Sham potential $\partial_{\mathbf{q}; \kappa \alpha} V(\mathbf{r})$ is
decomposed as
\begin{equation}
    \partial_{\mathbf{q}; \kappa \alpha} V(\mathbf{r}) =
    \partial_{\mathbf{q}; \kappa \alpha} V^{\mathrm{S}}(\mathbf{r}) +
    \partial_{\mathbf{q}; \kappa \alpha} V^{\mathrm{L}}(\mathbf{r}).
    \label{eq:dVscf_LR_SR}
\end{equation}
The SR components are localized in real space and can therefore be smoothly
Fourier interpolated onto arbitrary $\mathbf{q}$ grids, whereas the LR
contributions are treated via analytical models that correctly reproduce the
non-analytic behavior in the $\mathbf{q} \to \mathbf{0}$ limit. For
three-dimensional systems, the LR component of the dynamical matrix is given
by~\cite{PhysRevB.55.10355}
\begin{equation}
    D^{\mathrm{L},\text{3D}}_{\kappa\alpha,\kappa'\beta}(\mathbf{q}) =
    \frac{4\pi e^2}{\Omega \sqrt{M_\kappa M_{\kappa'}}}
    \sum_{\mathbf{G}\ne -\mathbf{q}}
    \frac{
        \mathcal{M}_{\kappa\alpha}^*(\mathbf{q}+\mathbf{G})\,
        \mathcal{M}_{\kappa'\beta}(\mathbf{q}+\mathbf{G})
    }{
        (\mathbf{q}+\mathbf{G}) \cdot \boldsymbol{\epsilon}^\infty
        \cdot (\mathbf{q}+\mathbf{G})
    }
    \, e^{-i(\mathbf{q}+\mathbf{G})\cdot(\boldsymbol{\tau}_{\kappa'}-\boldsymbol{\tau}_\kappa)}
    \, e^{-\frac{(\mathbf{q}+\mathbf{G})\cdot\boldsymbol{\epsilon}^\infty
        \cdot(\mathbf{q}+\mathbf{G})}{4\eta}},
    \label{eq:dyn_LR_3D}
\end{equation}
and the LR component of the perturbed Kohn-Sham potential
is written as~\cite{PhysRevLett.115.176401,PhysRevLett.125.136601,
PhysRevB.102.094308,PhysRevB.102.125203,PhysRevLett.125.136602}
\begin{equation}
    \partial_{\mathbf{q};\kappa\alpha} V^{\mathrm{L},\text{3D}}(\mathbf{r}) =
    \frac{4\pi e^2}{\Omega}
    \sum_{\mathbf{G}\ne -\mathbf{q}}
    \frac{i\,\mathcal{M}_{\kappa\alpha}(\mathbf{q}+\mathbf{G})}{
        (\mathbf{q}+\mathbf{G})\cdot\boldsymbol{\epsilon}^\infty
        \cdot(\mathbf{q}+\mathbf{G})
    }
    \, e^{i(\mathbf{q}+\mathbf{G})\cdot(\mathbf{r}-\boldsymbol{\tau}_\kappa)}
    \, e^{-\frac{(\mathbf{q}+\mathbf{G})\cdot\boldsymbol{\epsilon}^\infty
        \cdot(\mathbf{q}+\mathbf{G})}{4\eta}}.
    \label{eq:dVscf_LR_3D}
\end{equation}
Here $\Omega$ is the unit-cell volume, $\mathbf{G}$ are reciprocal-lattice
vectors, $M_\kappa$ is the mass of atom $\kappa$ located at
$\boldsymbol{\tau}_\kappa$, and $\boldsymbol{\epsilon}^\infty$ is the
$3\times 3$ macroscopic dielectric tensor. The Ewald summation parameter is
denoted by $\eta$, and $\mathcal{M}_{\kappa\alpha}$ is
defined as
\begin{equation}
    \mathcal{M}_{\kappa\alpha}(\mathbf{q}) =
    \mathbf{q}\cdot\mathbf{Z}_{\kappa\alpha}
    - \frac{i}{2}\,\mathbf{q}\cdot\mathbf{Q}_{\kappa\alpha}\cdot\mathbf{q},
    \label{eq:vertex_3D}
\end{equation}
where $\mathbf{Z}$ and $\mathbf{Q}$ are the Born effective-charge and
dynamical quadrupole tensors, respectively.

For two-dimensional systems, the LR component of the dynamical matrix takes
the form~\cite{Sohier2017Jun,PhysRevX.11.041027}
\begin{align}
    D^{\mathrm{L},\text{2D}}_{\kappa\alpha,\kappa'\beta}(\mathbf{q}_{\parallel})
    &= \frac{2\pi e^2}{S\sqrt{M_\kappa M_{\kappa'}}}
    \sum_{\mathbf{G}_{\parallel}\ne -\mathbf{q}_{\parallel}}
    \frac{f(|\mathbf{q}_{\parallel}+\mathbf{G}_{\parallel}|)}{
          |\mathbf{q}_{\parallel}+\mathbf{G}_{\parallel}|}
    \nonumber\\
    &\quad\times
    \left[
        \frac{
            \mathcal{M}^{\parallel*}_{\kappa\alpha}
            (\mathbf{q}_{\parallel}+\mathbf{G}_{\parallel})\,
            \mathcal{M}^{\parallel}_{\kappa'\beta}
            (\mathbf{q}_{\parallel}+\mathbf{G}_{\parallel})
        }{
            \tilde{\epsilon}^{\parallel}
            (\mathbf{q}_{\parallel}+\mathbf{G}_{\parallel})
        }
        -
        \frac{
            \mathcal{M}^{\perp*}_{\kappa\alpha}
            (\mathbf{q}_{\parallel}+\mathbf{G}_{\parallel})\,
            \mathcal{M}^{\perp}_{\kappa'\beta}
            (\mathbf{q}_{\parallel}+\mathbf{G}_{\parallel})
        }{
            \tilde{\epsilon}^{\perp}
            (\mathbf{q}_{\parallel}+\mathbf{G}_{\parallel})
        }
    \right]
    e^{-i(\mathbf{q}_{\parallel}+\mathbf{G}_{\parallel})
       \cdot(\boldsymbol{\tau}_{\kappa'}-\boldsymbol{\tau}_\kappa)},
    \label{eq:dyn_LR_2D_DD}
\end{align}
where $S$ is the unit-cell area and
$f(|\mathbf{q}_{\parallel}|) = 1 -
\tanh\!\left(\tfrac{|\mathbf{q}_{\parallel}|L}{2}\right)$, with $L$ being
the range-separation parameter and 
$\mathcal{M}^{\parallel}_{\kappa\alpha}$,
$\mathcal{M}^{\perp}_{\kappa\alpha}$ are defined as
\begin{align}
    \mathcal{M}^{\parallel}_{\kappa\alpha}(\mathbf{q}_\parallel)
    &= \mathbf{q}_\parallel\cdot\mathbf{Z}_{\kappa\alpha}
       - \frac{i}{2}\,\mathbf{q}_\parallel\cdot
         \left(\hat{\mathbf{Q}}_{\kappa\alpha}
               - 
               \hat{{Q}}^{(zz)}_{\kappa\alpha}\,\mathbf{I}\right)
         \cdot\mathbf{q}_\parallel,
    \label{eq:M_par}\\
    \mathcal{M}^{\perp}_{\kappa\alpha}(\mathbf{q}_\parallel)
    &= |\mathbf{q}_\parallel|
       \left(
           \frac{\hat{z}\cdot\mathbf{Z}_{\kappa\alpha}}{\epsilon^\infty_{zz}}
           - i\,(\mathbf{q}_\parallel\cdot
           \hat{\mathbf{Q}}_{\kappa\alpha})^{(z)}
       \right),
    \label{eq:M_perp}
\end{align}
where $\hat{\mathbf{Q}}$ is the effective two-dimensional quadrupole tensor
defined as~\cite{PhysRevX.11.041027}
\begin{align}
    \hat{Q}^{(z\alpha)}_{\kappa\beta}
    &= \frac{Q^{(z\alpha)}_{\kappa\beta}
           + \tau_{\kappa z}\,Z^{(\alpha)}_{\kappa\beta}}
          {\epsilon^\infty_{zz}},
    \label{eq:Qhat_za}\\
    \hat{Q}^{(zz)}_{\kappa\beta}
    &= \frac{Q^{(zz)}_{\kappa\beta}
           + 2\tau_{\kappa z}\,Z^{(z)}_{\kappa\beta}}
          {\epsilon^\infty_{zz}},
    \label{eq:Qhat_zz}\\
    \hat{Q}^{(\alpha\gamma)}_{\kappa\beta}
    - \delta_{\alpha\gamma}\,\hat{Q}^{(zz)}_{\kappa\beta}
    &= Q^{(\alpha\gamma)}_{\kappa\beta}
    - \epsilon^\infty_{\alpha\gamma}\,\hat{Q}^{(zz)}_{\kappa\beta},
    \label{eq:Qhat_inplane}
\end{align}
where $\alpha, \gamma \in \{x,y\}$. The two-dimensional screening functions entering Eq.~\eqref{eq:dyn_LR_2D_DD} are given by~\cite{Sohier2017Jun,PhysRevX.11.041027}
\begin{align}
    \tilde{\epsilon}^{\parallel}(\mathbf{q}_{\parallel})
    &= 1 + \frac{2\pi f(|\mathbf{q}_{\parallel}|)}{|\mathbf{q}_{\parallel}|}
       \,\mathbf{q}_{\parallel}\cdot\boldsymbol{\alpha}_{\parallel}
       \cdot\mathbf{q}_{\parallel},
    \label{eq:eps_par}\\
    \tilde{\epsilon}^{\perp}(\mathbf{q}_{\parallel})
    &= 1 - 2\pi|\mathbf{q}_{\parallel}|\,f(|\mathbf{q}_{\parallel}|)\,\alpha_\perp,
    \label{eq:eps_perp}
\end{align}
with the two-dimensional polarizabilities
$\alpha_{\parallel}^{\alpha\beta} =
\tfrac{c}{4\pi}\!\left(\epsilon^\infty_{\alpha\beta}-\delta_{\alpha\beta}\right)$
and
$\alpha_\perp =
\tfrac{c}{4\pi}\!\left(1-(\epsilon^\infty_{zz})^{-1}\right)$,
where $c$ is the out-of-plane lattice constant. The corresponding LR component
of the perturbed Kohn-Sham potential
is written as~\cite{PhysRevX.11.041027,PhysRevB.106.115423,PhysRevB.107.155424}
\begin{equation}
    \partial_{\mathbf{q}_{\parallel};\kappa\alpha}
    V^{\mathrm{L},\text{2D}}(\mathbf{r}) =
    \frac{2i\pi e^2}{S}
    \sum_{\mathbf{G}_{\parallel}\ne-\mathbf{q}_{\parallel}}
    \frac{f(|\mathbf{q}_{\parallel}+\mathbf{G}_{\parallel}|)}{
          |\mathbf{q}_{\parallel}+\mathbf{G}_{\parallel}|}
    \left[
        \frac{
            \mathcal{M}^{\parallel}_{\kappa\alpha}
            (\mathbf{q}_{\parallel}+\mathbf{G}_{\parallel})
        }{
            \tilde{\epsilon}^{\parallel}
            (\mathbf{q}_{\parallel}+\mathbf{G}_{\parallel})
        }
        - i\,z |\mathbf{q}_{\parallel}+\mathbf{G}_{\parallel}|\,
        \frac{
            \mathcal{M}^{\perp}_{\kappa\alpha}
            (\mathbf{q}_{\parallel}+\mathbf{G}_{\parallel})
        }{
            \tilde{\epsilon}^{\perp}
            (\mathbf{q}_{\parallel}+\mathbf{G}_{\parallel})
        }
    \right]
    e^{i(\mathbf{q}_{\parallel}+\mathbf{G}_{\parallel})
       \cdot(\mathbf{r}-\boldsymbol{\tau}_\kappa)}.
    \label{eq:dVscf_LR_2D}
\end{equation}

We would like to note a few implementation details. For
three-dimensional systems, the default Ewald summation parameter is set to
$\eta = 1$ bohr$^{-2}$. For two-dimensional systems, the range-separation parameter $L$
must satisfy $L > L_{\min} = 4\pi\alpha_\perp$~\cite{PhysRevX.11.041027} and we adopt the default value $L = 1.01\,L_{\min}$. Although both parameters
should, in principle, be validated through convergence tests, the default
choices were found to provide satisfactory accuracy across all systems
considered in this work. Another important aspect is the numerical evaluation of
Eq.~\eqref{eq:dVscf_LR_2D}. Since standard plane-wave implementations rely
on three-dimensional FFTs, this expression must be reformulated in terms of a
three-dimensional Fourier representation. We achieve this via two
approximations: the inverse out-of-plane screening function is expanded as
\begin{equation}
    \frac{1}{\tilde{\epsilon}^{\perp}(\mathbf{q}_{\parallel})}
    \approx
    1 + 2\pi|\mathbf{q}_{\parallel}|\,f(|\mathbf{q}_{\parallel}|)\,\alpha_\perp
    + \mathcal{O}(|\mathbf{q}_{\parallel}|^2),
    \label{eq:eps_perp_expand}
\end{equation}
after which all $\mathcal{O}(|\mathbf{q}_{\parallel}|^2)$ contributions from
the out-of-plane term are discarded; and the real-space factors $z$ and $1$
multiplying $\mathcal{M}^{\perp}$ and $\mathcal{M}^{\parallel}$,
respectively, are expanded in a Fourier series following the Coulomb cutoff
approach of Ref.~\cite{PhysRevB.96.075448},
\begin{align}
    z &\;\longrightarrow\;
    \sum_{G_z\neq 0} \frac{i}{G_z}\cos\!\left(\frac{G_z c}{2}\right)
    e^{iG_z z},
    \label{eq:z_replacementz}\\
    1 &\;\longrightarrow\;
    \sum_{G_z} \delta_{G_z,0}\, e^{iG_z z},
    \label{eq:z_replacement1}
\end{align}
where $\delta_{G_z,0}$ is the Kronecker delta. Substituting
Eqs.~\eqref{eq:z_replacementz} and~\eqref{eq:z_replacement1} into
Eq.~\eqref{eq:dVscf_LR_2D} yields an expression that depends on the full
three-dimensional wavevector $\mathbf{G}$ and can therefore be
evaluated directly with standard 3D FFT routines.

Having obtained the LR analytical models and the SR components on the coarse
$\mathbf{q}$-point grid, the SR components of the dynamical matrix and
the perturbed Kohn-Sham potential are Fourier transformed to real space~\cite{PhysRevB.78.045124,PhysRevB.55.10355},
\begin{align}
    D^{\mathrm{S}}_{\kappa \alpha, \kappa' \alpha'}(\mathbf{R})
    &= \frac{1}{N_\mathbf{q}}\sum_{\mathbf{q}}
    e^{-i\mathbf{q}\cdot\mathbf{R}}
    D^{\mathrm{S}}_{\kappa \alpha, \kappa' \alpha'}(\mathbf{q}),
    \label{eq:Dyn_SR_R}\\
    \partial_{\mathbf{R}; \kappa \alpha} V^{\mathrm{S}}(\mathbf{r})
    &= \frac{1}{N_\mathbf{q}}\sum_{\mathbf{q}}
    e^{-i\mathbf{q}\cdot\mathbf{R}}
    \partial_{\mathbf{q}; \kappa \alpha} V^{\mathrm{S}}(\mathbf{r}),
    \label{eq:dVscf_SR_R}
\end{align}
where $N_\mathbf{q}$ is the total number of coarse $\mathbf{q}$-points and
$\mathbf{R}$ labels the coarse real-space grid. The dynamical matrix and
perturbed Kohn-Sham potential on an arbitrary fine $\mathbf{q}_f$-point grid
are then recovered by
\begin{align}
    D_{\kappa \alpha, \kappa' \alpha'}(\mathbf{q}_f)
    &= D^{\mathrm{L}}_{\kappa \alpha, \kappa' \alpha'}(\mathbf{q}_f)
    + \sum_{\mathbf{R}}
    D^{\mathrm{S}}_{\kappa \alpha, \kappa' \alpha'}(\mathbf{R})\,
    \frac{1}{N_{\text{degen}}}\sum_{\mathbf{T}}
    e^{i\mathbf{q}_f\cdot(\mathbf{R}+\mathbf{T})},
    \label{eq:Dyn_fine}\\
    \partial_{\mathbf{q}_f; \kappa \alpha} V(\mathbf{r})
    &= \partial_{\mathbf{q}_f; \kappa \alpha} V^{\mathrm{L}}(\mathbf{r})
    + \sum_{\mathbf{R}}
    \partial_{\mathbf{R}; \kappa \alpha} V^{\mathrm{S}}(\mathbf{r})\,
    \frac{1}{N_{\text{degen}}}\sum_{\mathbf{T}}
    e^{i\mathbf{q}_f\cdot(\mathbf{R}+\mathbf{T})}.
    \label{eq:dvscf_fine}
\end{align}
It is important to note the occurrence of $\mathbf{T}$, which is an integer multiple of the Born-von~K\'{a}rm\'{a}n
supercell lattice vectors in Eqs.~\eqref{eq:Dyn_fine} and~\eqref{eq:dvscf_fine}. Since the SR components are periodic under lattice translations of the
Born-von~K\'{a}rm\'{a}n supercell associated with the coarse grid,
\begin{equation}
    D^{\mathrm{S}}_{\kappa \alpha, \kappa' \alpha'}(\mathbf{R})
    = D^{\mathrm{S}}_{\kappa \alpha, \kappa' \alpha'}(\mathbf{R}+\mathbf{T}),
    \qquad
    \partial_{\mathbf{R}; \kappa \alpha} V^{\mathrm{S}}(\mathbf{r})
    = \partial_{\mathbf{R}+\mathbf{T}; \kappa \alpha} V^{\mathrm{S}}(\mathbf{r}),
    \label{eq:SR_periodicity}
\end{equation}
there is freedom in the choice of the Born-von~K\'{a}rm\'{a}n supercell in
which $\mathbf{R}$ resides. Following
Refs.~\cite{PhysRevLett.78.4063,Pizzi2020Jan}, we choose $\mathbf{T}$ such
that $\mathbf{T}$ minimizes
$|\boldsymbol{\tau}_{\kappa'} - \boldsymbol{\tau}_{\kappa} + \mathbf{R} +
\mathbf{T}|$, while for the perturbed potential, $\mathbf{T}$ minimizes
$|\mathbf{R}+\mathbf{T}|$. When multiple equivalent $\mathbf{T}$
vectors exist, all $N_{\text{degen}}$ degenerate images are included, and their contributions are 
averaged. In \texttt{LetzElPhC}, we employ a KD-tree-based 
binary search, which significantly accelerates the finding of the $\mathbf{T}$ vectors when performing interpolation.

In Fig.~\ref{fig:interpolation}, we compare the Brillouin-zone-averaged electron–phonon matrix elements computed using $\partial^{\nu}_{\mathbf{q}} V_{\sigma \sigma'}$ obtained directly from DFPT calculations with those obtained using the interpolation method. As shown in the figure, the interpolated matrix elements for the different phonon branches are in excellent agreement with the corresponding DFPT results across a range of systems, including bulk non-polar materials such as silicon, polar materials such as GaN, two-dimensional systems such as InSe, and magnetic systems such as bcc Fe. For the polar systems (GaN and InSe), we include both dipole and quadrupole contributions in the treatment of the long-range component, whereas for bulk Si only the quadrupole contribution is present. The dipole and quadrupole tensors are computed using the \texttt{Quantum ESPRESSO} code~\cite{PhysRevB.110.094306,hc53-g1p3}. The computational details of these calculations are summarized in Table~\ref{tab:interp_comp_details}. All calculations were performed using scalar-relativistic pseudopotentials from the SPMS, SG15, and PseudoDojo projects~\cite{PhysRevB.88.085117,vanSetten2018May,Shojaei2023Feb}.

\subsection{Acoustic Sum Rules}
\label{sec:asr}
The translational and rotational invariance of the crystal potential energy imposes constraints on the interatomic force constants (IFCs) ~\cite{hiphive,Born1998Sep,Lin2022Nov}. In practice, IFCs obtained from \textit{ab initio} calculations may slightly violate these constraints due to the finite size of the supercell ($\mathbf{q}$-point grid) or basis set, leading to spurious non-zero or imaginary frequencies for the acoustic modes at the $\Gamma$ point. It is therefore common practice to apply a minimal correction to the calculated IFCs $\Phi_{\kappa\alpha,\kappa'\beta}$ (where $\kappa$ and $\kappa'$ denote atomic indices and $\alpha$ and $\beta$ are Cartesian directions) to enforce the following constraints ~\cite{hiphive,Born1998Sep,Lin2022Nov}:
\begin{align}
    \sum_{\kappa'} \Phi_{\kappa\alpha,\kappa'\beta} &= 0
    \quad \text{(translational invariance)},
    \label{eq:asr_eq}\\
    \sum_{\kappa'} \left(
        \Phi_{\kappa\alpha,\kappa'\beta}\,\tau_{\kappa'\gamma}
        - \Phi_{\kappa\alpha,\kappa'\gamma}\,\tau_{\kappa'\beta}
    \right) &= 0
    \quad \text{(rotational invariance)},
    \label{eq:Born_Huang}\\
    \sum_{\kappa\kappa'} \left(
        \Phi_{\kappa\alpha,\kappa'\beta}\,\tau_{\kappa\kappa'\gamma}
        \tau_{\kappa\kappa'\delta}
        - \Phi_{\kappa\gamma,\kappa'\delta}\,\tau_{\kappa\kappa'\alpha}
        \tau_{\kappa\kappa'\beta}
    \right) &= 0
    \quad \text{(vanishing stress at equilibrium)},
    \label{eq:Huang}
\end{align}
where $\tau_{\kappa\gamma}$ is the equilibrium position of atom $\kappa$ along
the Cartesian direction $\gamma$ and
$\tau_{\kappa\kappa'\gamma} = \tau_{\kappa'\gamma} - \tau_{\kappa\gamma}$.
Enforcing these conditions on the \textit{ab initio} IFCs corresponds to finding the corrected force constants $\boldsymbol{\Phi}_{\mathrm{corr}}$ that satisfy $A\boldsymbol{\Phi}_{\mathrm{corr}}=\mathbf{0}$ while minimizing the deviation from the original IFCs, \begin{equation} \min_{\boldsymbol{\Phi}_{\mathrm{corr}}} \left\| \boldsymbol{\Phi}_{\mathrm{corr}} -\boldsymbol{\Phi}_{\mathrm{orig}} \right\|_2 . \end{equation} 
By writing the correction as a projection from the row space of $A$, $\boldsymbol{\Phi}_{\mathrm{corr}} = \boldsymbol{\Phi}_{\mathrm{orig}} - A^{T}\boldsymbol{\lambda}$, the problem reduces to finding the Lagrange multipliers $\boldsymbol{\lambda}$ that satisfy the normal equations $AA^{T}\boldsymbol{\lambda} = A\boldsymbol{\Phi}_{\mathrm{orig}}$, which is equivalent to solving the least-squares problem $\min_{\boldsymbol{\lambda}} \Vert A^T\boldsymbol{\lambda} - \boldsymbol{\Phi}_{\mathrm{orig}}\Vert_2$. Since explicitly storing and factorizing $A$ is impractical for large supercells, \texttt{LetzElPhC} solves this least-squares problem using the matrix-free LSMR iterative solver~\cite{Fong2011Oct}, which evaluates the required matrix-vector products implicitly without constructing the constraint matrix $A$ explicitly. This approach is both computationally and memory efficient, scalable to large supercells, and numerically robust for systems of varying sizes.
\begin{figure}
    \centering
    \includegraphics[width=0.5\linewidth]{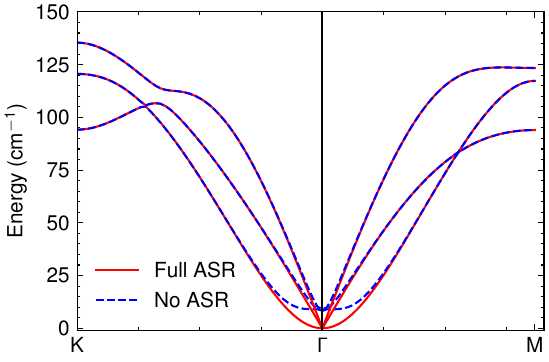}
    \caption{Dispersion relation of the acoustic phonons in monolayer WSe$_2$ computed with
and without enforcing the full sum rules using \texttt{LetzElPhC}.}
    \label{fig:wse2_asr}
\end{figure}

To demonstrate the implementation of the sum rules in the \texttt{LetzElPhC} code, we show the acoustic phonon modes of monolayer WSe$_2$ near the $\Gamma$ point in Fig.~\ref{fig:wse2_asr}. To obtain the phonon dispersion, we performed DFPT calculations on a uniform, $\Gamma$-centered $6\times6\times1$ $\mathbf{q}$-point grid and subsequently performed Fourier interpolation to obtain the dispersion curves. As shown in the figure, the blue dashed line corresponds to the dispersion obtained without correcting the calculated force constants. In this case, the force constants violate the acoustic sum rules, resulting in a finite frequency at the $\Gamma$ point. In addition, the parabolic nature of the out-of-plane acoustic mode is not accurately reproduced. In contrast, enforcing the sum rules on the force constants yields a phonon dispersion in which the acoustic modes have zero frequency at the $\Gamma$ point, while the out-of-plane mode exhibits the expected parabolic dispersion.

\section{Usage, Parallelization, and Test Suite}

\subsection{Usage}
\label{sec:usage}

The \texttt{LetzElPhC} package is executed through a unified command-line executable, \texttt{lelphc}, which provides three primary operating modes designed to manage each stage of the electron-phonon workflow:
\begin{enumerate}
    \item \textbf{Preprocessor mode} (\texttt{-pp}): parses and consolidates raw DFPT output files into a standardized, portable phonon directory;
    \item \textbf{Electron-phonon mode} (default): evaluates electron-phonon matrix elements and electronic symmetry representation matrices across the Brillouin zone;
    \item \textbf{Fourier interpolation mode} (\texttt{-i} or \texttt{--interpolate}): performs Fourier interpolation of dynamical matrices and perturbed Kohn-Sham potentials from coarse to fine momentum grids.
\end{enumerate}

Prior to running \texttt{LetzElPhC}, the required ground-state electronic and vibrational datasets must be generated using standard DFT and DFPT calculations. Specifically, one-electron Kohn-Sham wave functions and energies are calculated with \texttt{Quantum ESPRESSO} (\texttt{pw.x}) on an unshifted, uniform $\mathbf{k}$-point grid and converted into the \texttt{Yambo} database format using the \texttt{p2y} interface, which generates the \texttt{SAVE} directory. Phonon dynamical matrices and the variation of the self-consistent Kohn-Sham potential $\partial_{\mathbf{q}} V_{\mathrm{scf}}(\mathbf{r})$ are computed on a commensurate, uniform $\Gamma$-centered $\mathbf{q}$-point grid using \texttt{ph.x}.

\subsubsection{Preprocessor mode}

In standard \texttt{ph.x} calculations, dynamical matrices, variations of the self-consistent potential, and response tensors, such as Born effective charges and the dielectric matrix, are distributed across internal scratch subdirectories (e.g., \texttt{\_ph0/}). The preprocessor mode streamlines data management by extracting the relevant quantities and collecting them in a self-contained directory named \texttt{ph\_save}, or in a custom directory specified through the \texttt{ELPH\_PH\_SAVE\_DIR} environment variable. The preprocessor is invoked from the command line as follows:
\begin{center}
\begin{minipage}{\textwidth}
\centering
\texttt{lelphc -pp --code=qe -F ph.x\_input\_file}
\end{minipage}
\end{center}

Here, \texttt{ph.x\_input\_file} denotes the input file used for the corresponding \texttt{ph.x} calculation. The preprocessor parses this file to identify the calculation prefix and scratch directory and extracts the quantities required for the subsequent electron-phonon calculations. 

\subsubsection{Electron-phonon mode}
The electron-phonon mode represents the primary execution mode of \texttt{LetzElPhC}. It computes the electron-phonon matrix elements $\mathfrak{g}_{mn}^\nu(\mathbf{k},\mathbf{q})$ defined in Eq.~\eqref{eq:gkkp} alongside the electronic symmetry representation matrices $\mathcal{D}_{\mathbf{k}}(g)$ defined in Eq.~\eqref{eq:Dmat}. The calculation is executed in parallel via MPI using:
\begin{center}
\begin{minipage}{\textwidth}
\centering
\texttt{mpirun -n N lelphc -F lelphc.in}
\end{minipage}
\end{center}
where $N$ is the total number of MPI processes and \texttt{lelphc.in} is the input file. An example input file is shown in Listing~\ref{lst:elph_input}.

\begin{mylisting}[title={Listing~\thetcbcounter: Sample input file for electron-phonon calculations (\texttt{lelphc.in})},label={lst:elph_input}]
start_bnd       = 1
end_bnd         = 40
save_dir        = SAVE
ph_save_dir     = ph_save
kernel          = dfpt
convention      = standard
nkpool          = 2
nqpool          = 4
\end{mylisting}

The input parameters are defined as follows:
\begin{itemize}
    \item \texttt{start\_bnd} and \texttt{end\_bnd}: the 1-based indices defining the active band window $[n_{\min}, n_{\max}]$ of electronic Kohn-Sham states. 
    \item \texttt{save\_dir} and \texttt{ph\_save\_dir}: file paths pointing to the \texttt{Yambo} wave function database directory (default \texttt{SAVE}) and the preprocessed phonon directory (default \texttt{ph\_save}), respectively.
    \item \texttt{kernel}: specifies the level of physical screening included in the perturbing potential:
    \begin{itemize}
        \item \texttt{dfpt}: full self-consistent variation of the Kohn-Sham potential, including Hartree and exchange-correlation responses, local ionic variation, and non-local pseudopotential contributions;
        \item \texttt{dfpt\_local}: DFPT screening excluding non-local pseudopotential variations;
        \item \texttt{bare}: unscreened bare electron-ion potential variation including non-local pseudopotential terms;
        \item \texttt{bare\_local}: local component of the bare ionic potential variation only.
    \end{itemize}
    \item \texttt{convention}: specifies the momentum-transfer convention:
    \begin{itemize}
        \item \texttt{standard}: evaluates transition amplitudes according to $\langle m, \mathbf{k}+\mathbf{q} \mid \partial_{\mathbf{q}}^\nu V \mid n, \mathbf{k} \rangle$ [Eq.~\eqref{eq:gkkp}];
        \item \texttt{yambo}: evaluates transition amplitudes as $\langle m, \mathbf{k} \mid \partial_{\mathbf{q}}^\nu V \mid n, \mathbf{k}-\mathbf{q} \rangle$, matching the internal convention adopted by the \texttt{Yambo} code.
    \end{itemize}
    \item \texttt{nkpool} and \texttt{nqpool}: the number of MPI process pools assigned to parallelize over electron wavevectors $\mathbf{k}$ and phonon wavevectors $\mathbf{q}$, respectively (see Sec.~\ref{sec:paralleization}).
\end{itemize}

Upon completion, \texttt{LetzElPhC} produces two self-describing NetCDF-4 databases:
\begin{enumerate}
    \item \texttt{ndb.elph}: contains the unnormalized electron-phonon matrix elements $\mathfrak{g}_{mn}^\nu(\mathbf{k},\mathbf{q})$ (without $\sqrt{\frac{\hbar}{2\omega^\nu_{\mathbf{q}}}}$ factor), phonon frequencies $\omega_\mathbf{q}^\nu$, polarization displacement vectors $u_{\mathbf{q},\kappa\alpha}^\nu = M_\kappa^{-1/2} (\mathbf{d}^{\nu}_{\mathbf{q}})_{\kappa\alpha}$, and associated tensor metadata.
    \item \texttt{ndb.Dmats}: contains the electronic representation matrices $\mathcal{D}_{\mathbf{k},mn}(g)$ for each space-group symmetry operation $g \in \mathcal{G}$. These matrices are essential for applying the symmetry-unfolding relations in Eqs.~\eqref{eq:elph_rot222} and \eqref{eq:rot_exwf}, enabling the evaluation of exciton-phonon matrix elements in Eq.~\eqref{eq:exph_matrix} across the full Brillouin zone without phase ambiguities.
\end{enumerate}

\subsubsection{Fourier interpolation mode}
\texttt{LetzElPhC} provides a Fourier interpolation mode that interpolates dynamical matrices and perturbed Kohn-Sham potentials from coarse $\mathbf{q}$-meshes onto arbitrarily fine grids or paths, following the formalism outlined in Sec.~\ref{sec:interpolation}. The interpolation mode is triggered using:
\begin{center}
\begin{minipage}{\textwidth}
\centering
\texttt{lelphc -i --code=qe -F interpolate.in}
\end{minipage}
\end{center}
where \texttt{interpolate.in} is the input file controlling the interpolation parameters. A representative input file is shown in Listing~\ref{lst:interp_input}.

\begin{mylisting}[title={Listing~\thetcbcounter: Sample input file for Fourier interpolation (\texttt{interpolate.in})},label={lst:interp_input}]
ph_save_dir               = ph_save
ph_save_interpolation_dir = ph_save_interpolation
interpolate_dvscf         = true
nq1 = 24
nq2 = 24
nq3 = 24
asr                       = all_huang
zasr                      = crystal
loto                      = false
eta_ph                    = 1.0
eta_induced               = 1.0
write_dVbare              = false
\end{mylisting}

The input variables for interpolation include:
\begin{itemize}
    \item \texttt{ph\_save\_dir} and \texttt{ph\_save\_interpolation\_dir}: paths to the input coarse phonon directory and output interpolated phonon directory, respectively. The resulting interpolated directory shares the exact structure of a standard \texttt{ph\_save} directory and can be directly used as the \texttt{ph\_save\_dir} in subsequent fine-grid electron-phonon calculations.
    \item \texttt{interpolate\_dvscf}: logical switch indicating whether to interpolate the deformation potentials $\partial_{\mathbf{q}} V_{\mathrm{scf}}(\mathbf{r})$ alongside the dynamical matrices.
    \item \texttt{nq1}, \texttt{nq2}, \texttt{nq3}: dimensions of the fine uniform $\mathbf{q}$-point mesh. Alternatively, an explicit list of arbitrary crystal momenta (e.g., along high-symmetry band paths) can be provided via \texttt{qlist\_file}.
    \item \texttt{eta\_ph} and \texttt{eta\_induced}: Range-separation parameters for dynamical matrices and deformation potentials, respectively [Eqs.~\eqref{eq:dyn_LR_3D} and \eqref{eq:dVscf_LR_3D} in 3D, and Eqs.~\eqref{eq:dyn_LR_2D_DD} and \eqref{eq:dVscf_LR_2D} in 2D]. For two-dimensional materials, the effective slab thickness is automatically bounded by $L \ge 4\pi\alpha_\perp$ to ensure numerical stability.
    \item \texttt{asr} and \texttt{zasr}: specifies the acoustic sum rule applied to interatomic force constants and Born charges. Options for \texttt{asr} include \texttt{simple} (diagonal correction), \texttt{crystal} (translational invariance), \texttt{all} (translational and rotational invariance), and \texttt{all\_huang} (translational, rotational, and Huang stress invariances enforced via the iterative LSMR solver described in Sec.~\ref{sec:asr}).
    \item \texttt{loto} and \texttt{loto\_dir}: controls the non-analytic LO-TO splitting correction at the $\Gamma$ point along a designated limiting direction $\mathbf{q} \to \mathbf{0}$.
    \item \texttt{write\_dVbare}: optional logical flag to output the interpolated local bare potential variation $e^{-i\mathbf{q}\cdot\mathbf{r}}\partial_{\mathbf{q}}^\nu V_{\mathrm{loc}}(\mathbf{r})$ into an additional NetCDF database (\texttt{ndb.dVbare}).
\end{itemize}

\subsection{Parallelization}
\label{sec:paralleization}
\begin{figure}[tb]
\centering
\begin{tikzpicture}[
    >=Stealth,
    x=1.38cm, y=0.72cm,
    base/.style={rectangle, rounded corners=2.5pt, line width=0.5pt, align=center, inner sep=3pt, font=\scriptsize},
    rbox/.style={base, fill=blue!8, draw=blue!60!black, font=\scriptsize\bfseries},
    qbox/.style={base, fill=green!8, draw=green!50!black},
    kbox/.style={base, fill=orange!10, draw=orange!70!black},
    pwbox/.style={base, fill=purple!8, draw=purple!60!black},
    ompbox/.style={base, fill=red!8, draw=red!60!black},
    arr/.style={->, line width=0.5pt, draw=gray!70}
]

\node[font=\footnotesize\bfseries, text=blue!60!black] at (0, 3.6) {Total MPI};
\node[font=\footnotesize\bfseries, text=green!50!black] at (2.45, 3.6) {Level 1: $\mathbf{q}$-Pools};
\node[font=\footnotesize\bfseries, text=orange!70!black] at (4.90, 3.6) {Level 2: $\mathbf{k}$-Pools};
\node[font=\footnotesize\bfseries, text=purple!60!black] at (7.35, 3.6) {Level 3: Plane Waves};
\node[font=\footnotesize\bfseries, text=red!60!black] at (9.80, 3.6) {Level 4: OpenMP};

\node[rbox, text width=2.0cm] (root) at (0, 0) {12 CPUs\\ \texttt{nq}=2, \texttt{nk}=3};

\node[qbox, text width=2.1cm] (qp1) at (2.45, 1.8) {\textbf{q-Pool 1} (1--6)\\ Subset of $\mathbf{q}$-pts};
\node[qbox, text width=2.1cm] (qp2) at (2.45, -1.8) {\textbf{q-Pool 2} (7--12)\\ Subset of $\mathbf{q}$-pts};

\draw[arr] (root.east) -- ++(0.18,0) |- (qp1.west);
\draw[arr] (root.east) -- ++(0.18,0) |- (qp2.west);

\node[kbox, text width=2.0cm] (kp11) at (4.90, 2.7) {\textbf{k-Pool 1} (CPUs 1--2)};
\node[kbox, text width=2.0cm] (kp12) at (4.90, 1.8) {\textbf{k-Pool 2} (CPUs 3--4)};
\node[kbox, text width=2.0cm] (kp13) at (4.90, 0.9) {\textbf{k-Pool 3} (CPUs 5--6)};

\draw[arr] (qp1.east) -- ++(0.18,0) |- (kp11.west);
\draw[arr] (qp1.east) -- (kp12.west);
\draw[arr] (qp1.east) -- ++(0.18,0) |- (kp13.west);

\node[kbox, text width=2.0cm] (kp21) at (4.90, -0.9) {\textbf{k-Pool 1} (CPUs 7--8)};
\node[kbox, text width=2.0cm] (kp22) at (4.90, -1.8) {\textbf{k-Pool 2} (CPUs  9--10)};
\node[kbox, text width=2.0cm] (kp23) at (4.90, -2.7) {\textbf{k-Pool 3} (CPUs 11--12)};

\draw[arr] (qp2.east) -- ++(0.18,0) |- (kp21.west);
\draw[arr] (qp2.east) -- (kp22.west);
\draw[arr] (qp2.east) -- ++(0.18,0) |- (kp23.west);

\node[pwbox, text width=2.0cm] (pw11) at (7.35, 2.7) {FFT Slab (2 CPUs)};
\node[pwbox, text width=2.0cm] (pw12) at (7.35, 1.8) {FFT Slab (2 CPUs)};
\node[pwbox, text width=2.0cm] (pw13) at (7.35, 0.9) {FFT Slab (2 CPUs)};

\draw[arr] (kp11.east) -- (pw11.west);
\draw[arr] (kp12.east) -- (pw12.west);
\draw[arr] (kp13.east) -- (pw13.west);

\node[pwbox, text width=2.0cm] (pw21) at (7.35, -0.9) {FFT Slab (2 CPUs)};
\node[pwbox, text width=2.0cm] (pw22) at (7.35, -1.8) {FFT Slab (2 CPUs)};
\node[pwbox, text width=2.0cm] (pw23) at (7.35, -2.7) {FFT Slab (2 CPUs)};

\draw[arr] (kp21.east) -- (pw21.west);
\draw[arr] (kp22.east) -- (pw22.west);
\draw[arr] (kp23.east) -- (pw23.west);

\node[ompbox, text width=2.0cm] (omp11) at (9.80, 2.7) {Threading};
\node[ompbox, text width=2.0cm] (omp12) at (9.80, 1.8) {Threading};
\node[ompbox, text width=2.0cm] (omp13) at (9.80, 0.9) {Threading};

\draw[arr] (pw11.east) -- (omp11.west);
\draw[arr] (pw12.east) -- (omp12.west);
\draw[arr] (pw13.east) -- (omp13.west);

\node[ompbox, text width=2.0cm] (omp21) at (9.80, -0.9) {Threading};
\node[ompbox, text width=2.0cm] (omp22) at (9.80, -1.8) {Threading};
\node[ompbox, text width=2.0cm] (omp23) at (9.80, -2.7) {Threading};

\draw[arr] (pw21.east) -- (omp21.west);
\draw[arr] (pw22.east) -- (omp22.west);
\draw[arr] (pw23.east) -- (omp23.west);

\end{tikzpicture}
\caption{Hierarchical multi-level parallelization scheme in \texttt{LetzElPhC}, illustrated for 12 MPI processes partitioned with \texttt{nqpool = 2} (Level 1) and \texttt{nkpool = 3} (Level 2). At the lowest MPI level, the remaining processes parallelize over plane waves (Level 3), complemented by multithreaded OpenMP execution on each core (Level 4).}
\label{fig:parallel_tree}
\end{figure}
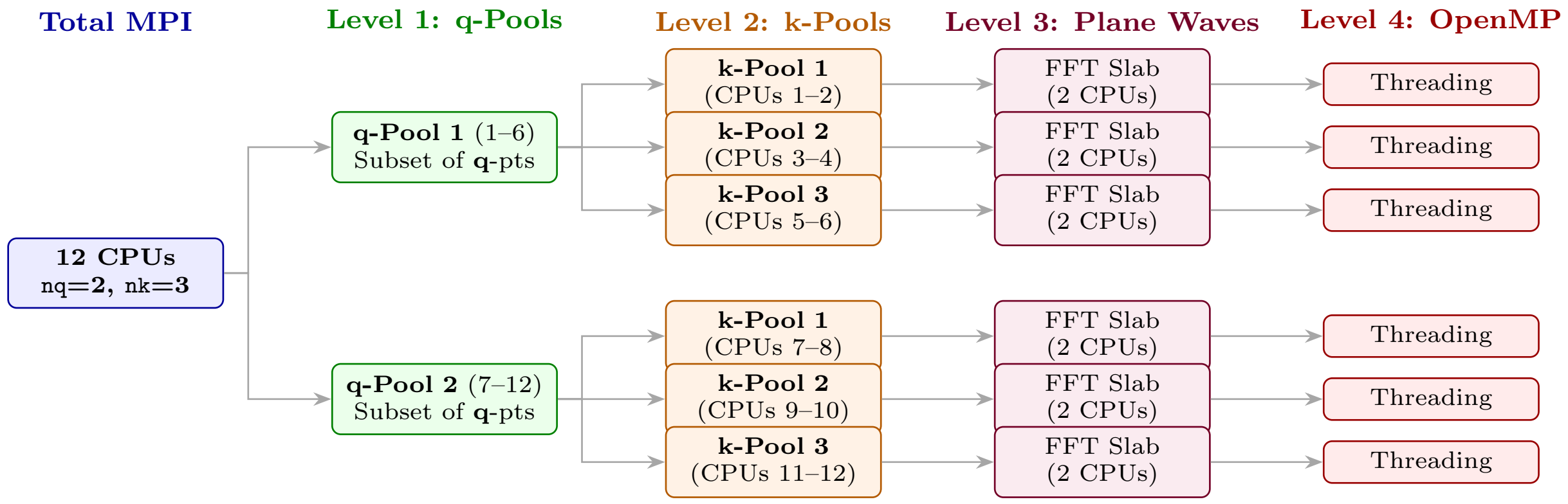

\begin{figure}
    \centering
    \includegraphics[width=0.5\linewidth]{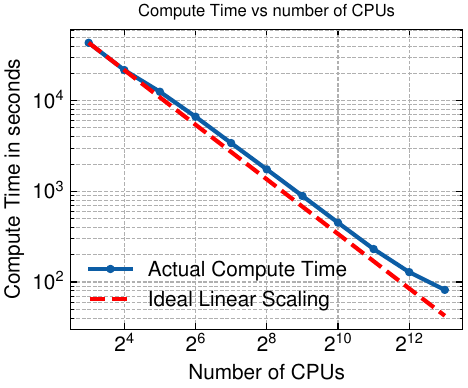}
    \caption{Strong-scaling performance for evaluating a $(2304,2304,9,8,8)$
electron-phonon matrix-element array in single precision, showing the total
computation time (excluding I/O time) as a function of the number of CPUs.}
    \label{fig:performace_scaling}
\end{figure}

The evaluation of electron-phonon matrix elements in converged exciton-phonon calculations requires efficient handling of large wave functions and perturbed-potential datasets on dense momentum grids. To address this computational challenge, \texttt{LetzElPhC} is implemented in C99 and designed for modern high-performance computing (HPC) environments. The code employs a hybrid MPI/OpenMP parallelization strategy, together with collective parallel I/O and distributed Fourier-transform algorithms, to distribute both the computational workload and memory requirements across a large number of processors.

The parallelization strategy has four levels as shown in Fig.~\ref{fig:parallel_tree}. At the first two levels, independent calculations are distributed over $\mathbf{q}$-points and $\mathbf{k}$-points using MPI. Since electron-phonon matrix elements must be evaluated for a large number of $\mathbf{q}$- and $\mathbf{k}$-points, these levels provide a high degree of parallelism. However, they do not reduce the memory required to process an individual $\mathbf{k}$-point, as the wave functions and perturbed potentials defined on reciprocal- and real-space grids can be too large to store on a single process.

This limitation is addressed at the third level, where the plane waves representing the wave functions and perturbed potentials are distributed across a group of MPI processes. Since evaluating the electron-phonon matrix elements requires Fourier transforms between reciprocal and real space, this level relies on a distributed three-dimensional FFT based on slab decomposition~\cite{10.1145/2063384.2063479}. The real-space data are distributed along the $z$ axis, while the reciprocal-space data are distributed across MPI processes as $(G_x,G_y)$ sticks along the $z$ direction. This approach avoids storing large intermediate arrays on individual processes and enables the use of the dense grids required for exciton-phonon calculations. Finally, at the fourth level, OpenMP threads within each MPI process accelerate computationally intensive operations while reducing communication overhead. Together, these four levels enable efficient utilization of modern HPC architectures while keeping the memory footprint per MPI process low.

In Fig.~\ref{fig:performace_scaling}, we show the strong-scaling performance of \texttt{LetzElPhC}, excluding I/O time. Specifically, we report the total compute time required to calculate the electron-phonon matrix elements of monolayer WSe$_2$, including spin-orbit coupling, using 4 conduction bands and 4 valence bands on uniform $\Gamma$-centered $48\times48\times1$ $\mathbf{k}$- and $\mathbf{q}$-point grids. The calculations were performed using between 8 and 8192 CPU processes, with 8 processes assigned to $\mathbf{G}$-space parallelization and the remaining processes distributed over the $\mathbf{q}$- and $\mathbf{k}$-point parallelization levels. As shown in the figure, \texttt{LetzElPhC} exhibits good strong scaling across the full range of process counts considered, demonstrating the efficient distribution of both the computational workload and memory requirements. At large process counts, the workload per process decreases, and MPI communication becomes increasingly significant, leading to deviations from ideal scaling.

The large size of the wave functions and perturbed-potential datasets also requires an efficient I/O strategy. \texttt{LetzElPhC} uses collective parallel I/O through the NetCDF4/HDF5 interface, which allows multiple MPI processes to access distributed datasets simultaneously. This avoids serial I/O bottlenecks and enables efficient handling of the large electron-phonon databases required for converged exciton-phonon calculations.

\subsection{Test suite}

To ensure the numerical reliability and reproducibility of the implementation,
\texttt{LetzElPhC} includes an automated test suite implemented in Python~3.
The framework uses \texttt{numpy}~\cite{Harris2020Sep}, \texttt{scipy}~\cite{Virtanen2020Mar}, and \texttt{netCDF4} libraries to
compare calculated quantities against reference results obtained from
validated calculations.

The DFPT datasets and reference databases required for the tests can be large
and are therefore not included directly in the main repository. Instead, they
are downloaded separately when initializing the test suite. For example, the
required reference datasets can be obtained using:
\begin{center}
\begin{minipage}{\textwidth}
\centering
\texttt{python3 run\_tests.py --download}
\end{minipage}
\end{center}

The complete test suite can then be executed by specifying the MPI launcher,
the number of processors, and the compiled \texttt{LetzElPhC} executable:
\begin{center}
\begin{minipage}{\textwidth}
\centering
\texttt{python3 run\_tests.py --ncpus=1 --mpirun=mpirun --lelphc=./lelphc}
\end{minipage}
\end{center}

Each test compares the generated NetCDF databases against the corresponding
reference databases within predefined numerical tolerances and verifies the
consistency of the results across different parallelization configurations.

\section{Applications}

\subsection{Resonant Raman scattering in monolayer WSe$_2$}

\begin{figure}
    \centering
    \includegraphics[width=0.5\linewidth]{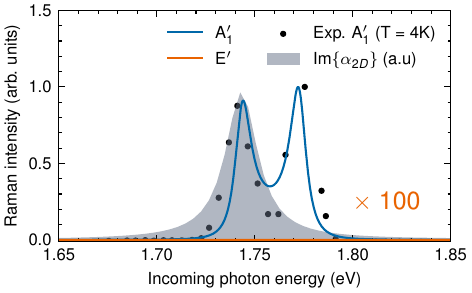}
    \caption{Resonant Raman intensities of the two Raman-active phonon modes of monolayer WSe$_2$ as a function of the laser excitation energy. The gray shading represents the absorption spectrum calculated using the GW-BSE formalism. The black dots denote experimental Raman data measured at 4~K, taken from Ref.~\cite{McDonnell_2020}. The calculated absorption spectrum and Raman peaks are rigidly shifted to match the experimental peak positions.}
    \label{fig:wse2_raman}
\end{figure}

As a first application, we consider resonant Raman scattering in monolayer WSe$_2$. Following Refs.~\cite{Reichardt2020Aug,PhysRevB.99.174312}, the one-phonon resonant Raman intensity, including excitonic effects, is given by
\begin{equation}
    \label{eq:tda_ram11}
    \begin{split}
        I^\lambda \propto \frac{\omega_{\mathrm{L}}-\omega_{\lambda}}{\omega_{\mathrm{L}}}
        \Bigg| \sum_{S, S^{\prime}} \frac{ \left(d^\nu_{S}\right)^* \left(\mathscr{G}_{S^\prime S}^\lambda(\mathbf{0},\mathbf{0})\right)^* d_{S^\prime}^\mu }{\left(\hbar \omega_{\mathrm{L}}-E_{S^\prime}+i \gamma\right)\left(\hbar \omega_{\mathrm{L}}-\hbar \omega_\lambda-E_{S}+i \gamma\right)}
         + \sum_{S, S^{\prime}} \frac{\left(d_{S^\prime}^\mu\right)^* \mathscr{G}_{S^\prime S}^\lambda (\mathbf{0},\mathbf{0})d^\nu_{S}}{\left(\hbar \omega_{\mathrm{L}}+E_{S^\prime}-i \gamma\right)\left(\hbar \omega_{\mathrm{L}}-\hbar \omega_\lambda+E_{S}-i \gamma\right)} \Bigg|^2,
    \end{split}
\end{equation}
where $d^\nu_S = \bra{S}\hat{v}^\nu\ket{0}$ denotes the exciton dipole matrix element for state $S$, with $\hat{v}$ being the velocity operator, and the exciton-phonon matrix element $\mathscr{G}^{\lambda}_{S'S}(\mathbf{0},\mathbf{0})$ defined in Eq.~\eqref{eq:exph_matrix}. Here, $\nu$ and $\mu$ denote the polarizations of the outgoing and incoming photons, respectively; $\gamma$ is the excitonic broadening parameter, $\omega_\lambda$ is the frequency of phonon mode $\lambda$ at crystal momentum $\mathbf{q}=\mathbf{0}$, $\omega_{\mathrm{L}}$ is the frequency of the incoming laser, and $E_S$ is the energy of the excitonic state $S$.

To evaluate Eq.~\eqref{eq:tda_ram11}, we first performed ground-state DFT and DFPT calculations using the \texttt{Quantum ESPRESSO} package with the PBE functional~\cite{PhysRevLett.77.3865} and fully relativistic optimized norm-conserving pseudopotentials from PseudoDojo~\cite{PhysRevB.88.085117,vanSetten2018May}. The plane-wave wave function cutoff was set to 120~Ry, and a $\Gamma$-centered $12\times 12\times 1$ $\mathbf{k}$-point grid was used to converge the ground-state density. A vacuum separation of 17~\AA{} was employed, together with a two-dimensional Coulomb cutoff to prevent spurious interactions between periodic images~\cite{PhysRevB.96.075448}. The electron-phonon matrix elements were computed using \texttt{LetzElPhC}, while the GW-BSE calculations were performed using \texttt{Yambo} on a $\Gamma$-centered $48\times 48\times 1$ $\mathbf{k}$-point grid, with the plasmon-pole approximation~\cite{PhysRevLett.62.1169}. In the GW step, the dielectric cutoff was set to 10~Ry, and 300 bands were included. The BSE Hamiltonian was constructed within the Tamm-Dancoff approximation~\cite{tda_ref}, including 2 valence and 2 conduction bands. Throughout the workflow, all calculations were performed in the irreducible Brillouin zone (IBZ), without breaking symmetries at any stage. We included 1000 states in each of the $S$ and $S'$ summations in Eq.~\eqref{eq:tda_ram11}.

Figure~\ref{fig:wse2_raman} shows the calculated resonant Raman intensities of the $A_1'$ (blue line) and $E'$ (orange line) phonon modes as functions of the incoming photon energy, along with the absorption spectrum (gray shading). We first compare the calculated $A_1'$ Raman intensity with experimental data measured at $T=4$~K~\cite{McDonnell_2020}, focusing on the energy range near the optical absorption onset, where the absorption is dominated by the $A_{1s}$ exciton. As the laser excitation energy is tuned across the $A_{1s}$ resonance, the $A_1'$ Raman intensity exhibits two pronounced peaks associated with intravalley scattering of $A_{1s}$ excitons. The first peak occurs at the $A_{1s}$ exciton energy, $E_{A_{1s}}$, while the second appears at $E_{A_{1s}}+\hbar\omega_{A_1'}$, shifted upwards by one $A_1'$ phonon energy. These peak positions are consistent with the poles of Eq.~\eqref{eq:tda_ram11} at $E_S$ and $E_S+\hbar\omega_\lambda$, respectively. The calculated peak structure agrees well with the experimental data.

The calculated Raman intensity of the $E'$ mode, by contrast, is several orders of magnitude weaker than that of the $A_1'$ mode, despite both modes being Raman active. This difference can be understood from the selection rules imposed by the threefold rotational symmetry at the $\Gamma$ point, which require the conservation of the total out-of-plane crystal angular momentum modulo 3~\cite{7h2w-ys9h}. The $A_1'$ phonon, with $j_z=0$, can mediate intravalley scattering, whereas the $E'$ phonon, with $j_z=\pm 1$, allows intervalley scattering between excitons in opposite $K$ and $K'$ valleys. The latter process is strongly suppressed by spin-selection rules and the negligible overlap between excitonic wave functions localized in opposite valleys. This suppression explains the much weaker resonant enhancement of the $E'$ mode compared with the $A_1'$ mode. These selection rules are also observed in monolayer MoSe$_2$ and in WSe$_2$/$h$BN heterostructures~\cite{Nalabothula2025Apr,7h2w-ys9h}.

\subsection{Phonon-assisted luminescence in face-centered cubic AgBr}

\begin{figure}
    \centering
    \includegraphics[width=\linewidth]{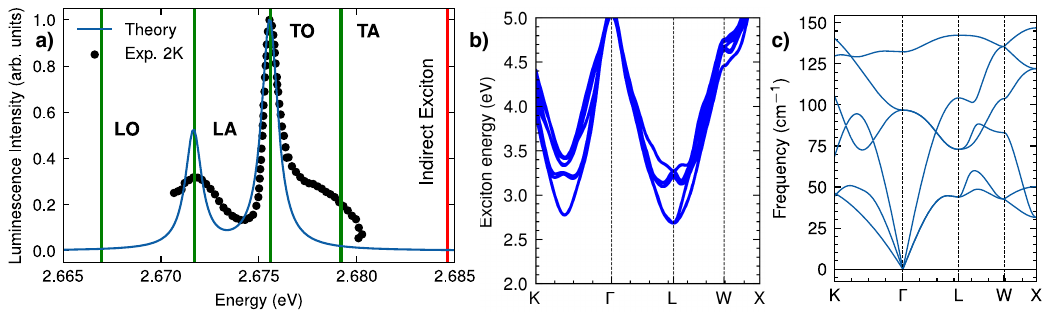}
    \caption{(a) Calculated and experimental phonon-assisted luminescence spectra of face-centered cubic AgBr. The experimental data are taken from Ref.~\cite{Kanzaki1970Oct}. The calculated spectrum is rigidly shifted by 1.06962~eV to match the experimental peak positions. The red line marks the position of the lowest indirect exciton, while the green vertical lines mark the positions of the phonon-assisted sidebands. (b) Interpolated exciton dispersion obtained from GW-BSE calculations, which is also rigidly shifted upward. (c) Phonon dispersion obtained from DFPT calculations. An interactive version of this figure, explaining the symmetries of the contributing phonons and electron-hole pair excitations, is available as supplementary material~\cite{suppli_info}, which includes Refs.~\cite{RevModPhys.40.1,Bradley1972,Koster1963,Altmann1994,Mulliken1955,Bouckaert1936,Giannozzi2009}.}
    \label{fig:agbr_lumin}
\end{figure}

As a second application, we consider phonon-assisted luminescence in face-centered cubic AgBr (rock-salt structure), which exhibits strong excitonic effects. Our first-principles calculations reveal that AgBr has an indirect optical gap, with a binding energy of $\sim$135~meV for the lowest-energy bright exciton. AgBr is therefore an excellent test system for studying exciton-phonon interactions, and here we examine its phonon-assisted emission spectrum. We calculate the phonon-assisted luminescence spectrum following the approach of Ref.~\cite{PhysRevLett.131.206902}, in which the intensity is expressed as
\begin{equation}
\begin{aligned}
    I \propto \!  \sum_{S',\mathbf{Q},\lambda,\mu} &  \left\{
   \frac{ e^{-\frac{E_{S'}^{\mathbf{Q}} - E^{\vphantom{\mathbf{Q}}}_m}{k_B T}} \, (1 + n_{\lambda, \mathbf{Q}}) \,
   \delta\!\left(\hbar\omega^{\vphantom{\mathbf{Q}}}_\mathrm{L} - E^\mathbf{Q}_{S'}+ \hbar \omega^{\vphantom{\mathbf{Q}}}_{\lambda, \mathbf{Q}}\right)}
   {E^\mathbf{Q}_{S'} - \hbar \omega^{\vphantom{\mathbf{Q}}}_{\lambda, \mathbf{Q}}} \right.
   \left.  \left| \sum_{S}
   \frac{ \left(d^\mu_{S} \, \mathscr{G}^{\vphantom{*}\lambda}_{S'S}(\mathbf{0},\mathbf{Q})\right)^*}
   {E^{\vphantom{\mathbf{Q}}}_{S} -E^\mathbf{Q}_{S'}+ \hbar \omega^{\vphantom{\mathbf{Q}}}_{\lambda, \mathbf{Q}}}
   \right|^2 \right\} ,
\end{aligned}
    \label{eq:ph_assited_lumin_expre}
\end{equation}
where $\omega_\mathrm{L}$ is the light emission frequency, $E_S^\mathbf{Q}$ denotes the energy of the exciton state $S$ with crystal momentum $\mathbf{Q}$, $n_{\lambda,\mathbf{Q}}$ is the Bose occupation factor for phonon mode $\lambda$ with crystal momentum $\mathbf{Q}$, $E_m$ is the energy of the lowest-energy exciton, and $T$ and $k_B$ denote the exciton temperature and the Boltzmann constant, respectively. The remaining indices follow the conventions of Eq.~\eqref{eq:tda_ram11}.

For AgBr, spin-orbit coupling was neglected in all calculations. The ground-state DFT calculation was performed within the local density approximation~\cite{PhysRev.136.B864}, using a plane-wave cutoff of 120~Ry and a $\Gamma$-centered $12\times 12\times 12$ $\mathbf{k}$-point grid. Phonon frequencies and eigenvectors were obtained from DFPT on a $12\times 12\times 12$ $\mathbf{q}$-point grid. The GW and BSE calculations were performed on the same uniform $\Gamma$-centered $12\times 12\times 12$ grid. The frequency dependence of the dielectric tensor was modeled using the plasmon-pole approximation~\cite{PhysRevLett.62.1169}. In both the GW and BSE steps, the dielectric cutoff was set to 15~Ry, and 150 bands were included in the calculation of the RPA screening and the GW self-energy. To obtain finite-momentum excitonic states across the $12\times 12\times 12$ $\mathbf{Q}$-point grid, the BSE Hamiltonian was constructed and solved using 3 valence bands and 1 conduction band. As in the monolayer WSe$_2$ example, all calculations were performed in the IBZ without breaking symmetries at any stage, and quantities throughout the full Brillouin zone were reconstructed using symmetry operations. We included 500 states in each of the $S$ and $S'$ summations in Eq.~\eqref{eq:ph_assited_lumin_expre}.

Figure~\ref{fig:agbr_lumin}a compares the calculated phonon-assisted luminescence spectrum of AgBr with experimental data~\cite{Kanzaki1970Oct}. The calculation reproduces the main spectral features, including the relative intensities and energy separations of the phonon replicas. To understand the origin of these features, we examine the exciton and phonon dispersions shown in Figs.~\ref{fig:agbr_lumin}b and \ref{fig:agbr_lumin}c, respectively. At thermal equilibrium, the Boltzmann factor in Eq.~\eqref{eq:ph_assited_lumin_expre} strongly favors excitons near the lowest exciton energy, $E_m$. At the exciton temperature of $T=2$~K, the population of higher-energy exciton states is exponentially suppressed. As shown in Fig.~\ref{fig:agbr_lumin}b, the lowest-energy exciton occurs at the $L$ point, $(0.5,0.5,0.5)$ in reduced reciprocal coordinates. The luminescence spectrum is therefore dominated by the phonon-assisted scattering of excitons near $L$ into optically bright excitonic states at $\Gamma$, involving phonons with crystal momentum $L$.

The phonon dispersion in Fig.~\ref{fig:agbr_lumin}c identifies the phonon modes that contribute to this process. Although there are four distinct energy modes at the $L$ point, only two replicas are observed in the calculated spectrum. The two dominant features correspond to the longitudinal acoustic (LA) and transverse optical (TO) modes. The absence of the highest energy (LO) and lowest energy (TA) phonon modes is due to the selection rules of the exciton-phonon interaction~\cite{7h2w-ys9h}. Since the dominant scattering path is from the lowest-energy excitons at the $L$ point to the bright excitons at the $\Gamma$ point, the phonon modes that can mediate this scattering are governed by the symmetries of these excitonic states. All bright excitons at $\Gamma$ transform as the $T_{1u}$ irrep of the $O_h$ little point group, which is odd under inversion. In contrast, the lowest-energy state at the $L$ point is doubly degenerate and transforms as the $E_g$ irrep of the $D_{3d}$ point group, which is even under inversion. This implies that the phonons mediating the exciton scattering from $ E_g \rightarrow T_{1u}$ must be odd under inversion in order to conserve parity. The LO and TA phonon modes at the $L$ point are even; therefore, these modes are forbidden and absent in the luminescence spectrum. This also emphasizes that our implementation correctly captures the selection rules, which are broken when gauge consistency is not maintained.

\section{Conclusion}

In summary, we have presented \texttt{LetzElPhC}, an open-source, high-performance C package designed to bridge density functional perturbation theory calculations in \texttt{Quantum ESPRESSO} with many-body excited-state workflows in the \texttt{Yambo} code. By reading one-electron Kohn-Sham wave functions directly from \texttt{Yambo} databases and applying a consistent symmetry-unfolding formalism to Bloch states, phonon eigenvectors, perturbed potentials, and excitonic wave functions, \texttt{LetzElPhC} guarantees strict phase and gauge consistency throughout the full Brillouin zone. This approach removes the need to disable crystal symmetries during electron-phonon and Bethe-Salpeter equation calculations, overcoming a major computational and memory bottleneck in first-principles exciton-phonon simulations. Moreover, this code can be easily ported to other plane-wave DFT codes and many-body perturbation theory codes.

To make calculations on dense momentum grids tractable for complex materials, \texttt{LetzElPhC} integrates Fourier interpolation of both dynamical matrices and perturbed Kohn-Sham potentials, accounting for long-range electrostatic dipole and dynamical quadrupole interactions in three- and two-dimensional crystals. Furthermore, the implementation of a matrix-free iterative solver allows translational, rotational, and equilibrium stress acoustic sum rules to be enforced efficiently on interatomic force constants. Combined with a four-level hybrid MPI/OpenMP parallelization hierarchy, distributed three-dimensional fast Fourier transforms, and efficient collective parallel I/O, the code scales from local workstations to thousands of CPU cores on modern high-performance computing architectures.

We have validated the interpolation accuracy and generality of \texttt{LetzElPhC} across bulk non-polar, polar, magnetic, and two-dimensional materials, and demonstrated its integration into \textit{ab initio} exciton-phonon workflows by computing resonant Raman scattering profiles in monolayer $\mathrm{WSe_2}$ and phonon-assisted luminescence spectra in face-centered cubic $\mathrm{AgBr}$. In both applications, our symmetry-preserving calculations reproduce experimental spectral features and selection rules at a fraction of the computational cost of full-Brillouin-zone evaluations. Together with its modular Python post-processing interface and automated test suite, \texttt{LetzElPhC} provides a scalable foundation for first-principles studies of phonon-assisted optical spectroscopies and exciton-phonon dynamics in emerging materials.

\section{Acknowledgements}
We acknowledge the use of the HPC facilities of the University of Luxembourg~\cite{VCPKVO_HPCCT22}. M.N. and L.W. acknowledge funding from the Luxembourg National Research Fund (FNR) through project C22/MS/17415967/ExcPhon. M.N. thanks Fulvio Paleari for the extensive testing of the code during its very early stages, Henry Fried for designing the code logo, and Sven Reichardt, Aseem Rajan Kshirsagar, Riccardo Reho, and Sahar Pakdel for helpful discussions and testing the code.

\section{Data Availability}
The data and software that support the findings of this article are openly available in the Zenodo repository~\cite{nalabothula_2026_24130523}. The code is publicly available on GitHub~\cite{letzcode_github}.

\bibliography{bibilo}
\end{document}